\documentstyle[11pt,aaspp4]{article}
\def	\beq	{\begin{equation}}
\def	\eeq	{\end{equation}}
\def	\Angstrom	{{\rm\AA}}
\def	\cm	{\,{\rm cm}}
\def	\etal	{et al.\ }
\def	\erg	{\,{\rm erg}}
\def	\eV	{\,{\rm eV}}
\def	\H	{{\rm H}}
\def	\He	{{\rm He}}
\def	\HH	{{\rm H}_2}
\def	\kms	{\,{\rm km\,s}^{-1}}
\def	\K	{\,{\rm K}}
\def	\mag	{\,{\rm mag}}
\def	\sol	{\odot}
\def	\pc	{\,{\rm pc}}
\def	\ltsim	{\lesssim}	%\lessim = aastex4 supplied macro
\def	\gtsim	{\gtrsim}	%\gtrsim = aastex4 supplied macro
\def	\nH	{n_{\rm H}}
\def	\NH	{N_{\rm H}}
\def	\s	{\,{\rm s}}
\def	\sr	{\,{\rm sr}}
\def	\hhini	{\delta}		%symbol for v-J distribution of newly formed H2
\lefthead{Draine \& Bertoldi}
\righthead{Stationary Photodissociation Fronts}

\begin{document}
\author{{\it Ap. J.}, in press (1996)}
\title{Structure of Stationary Photodissociation Fronts}
\author{B.T. Draine}
\affil{Princeton University Observatory, Peyton Hall,
	Princeton, NJ 08544, USA; {\tt draine@astro.princeton.edu}}

\and

\author{Frank Bertoldi}
\affil{Max-Planck-Institut f\"ur Extraterrestrische Physik,
	D-85748 Garching, Germany; {\tt fkb@mpe-garching.mpg.de}}
\begin{abstract}
The structure of stationary photodissociation fronts is revisited.
$\HH$ self-shielding is discussed, including
the effects of line overlap.
We find that line overlap is important for $N(\HH)\gtsim10^{20}\cm^{-2}$,
with a factor-of-two suppression of
pumping rates at column densities $N(\HH)\approx3\times10^{20}\cm^{-2}$.
We compute multiline UV pumping models, and compare these with simple
analytic approximations for the effects of self-shielding.

The overall fluorescent efficiency of the photodissociation front is
obtained for different ratios of $\chi/\nH$ (where $\chi$ characterizes
the intensity of the illuminating ultraviolet radiation) and different
dust extinction laws.
The dust optical depth $\tau_{pdr}$ to the point where 50\% of the H is
molecular is found to be a simple function of a dimensionless quantity
$\phi_0$ depending on $\chi/\nH$, the rate coefficient $R(T)$ for $\HH$
formation on grains, and the UV dust opacity.
The fluorescent efficiency of the PDR also depends primarily on $\phi_0$ for
$\chi\ltsim3000$ and $\nH\ltsim10^4\cm^{-3}$; for stronger radiation fields
and higher densities radiative and collisional depopulation of 
vibrationally-excited levels interferes with the radiative cascade.
We show that the emission spectrum from the PDR is essentially independent
of the color temperature $T_{color}$ of the illuminating radiation for
$10^4\K\ltsim T_{color}$, but shows some sensitivity to the rotation-vibration
distribution of newly-formed $\HH$.
The 1--0S(1)/2--1S(1) and 2--1S(1)/6--4Q(1) intensity ratios, 
the ortho/para ratio, and the
rotational temperature in the $v$=1 and $v$=2 levels are computed as
functions of the temperature and density, for different values of
$\chi/\nH$.

We apply our models to the reflection nebula NGC 2023.
Apparent inconsistencies between published K-band and
far-red spectroscopy of this object are discussed; 
we adjust the two sets of observations for consistency.
We are able to approximately
reproduce the (adjusted) observations with models having $\chi=5000$,
$\nH=10^5\cm^{-3}$, and a reasonable viewing angle.
Further observations of NGC 2023 will be valuable to clarify the
uncertain spatial structure of the emission.
\end{abstract}


\section{Introduction}

It is by now widely recognized that reprocessing of OB starlight in
photodissociation fronts plays an important role in the overall
energetics of star-forming galaxies, particularly extreme
``starburst'' galaxies.
An important mechanism in this reprocessing is the absorption of
ultraviolet photons by molecular hydrogen, followed by ultraviolet
and infrared fluorescence, and, about 15\% of the time,
by dissociation.
The vibrational fluorescence process,
first noted by Gould \& Harwit (1963), has been investigated
by a number of authors
(e.g.: Black \& Dalgarno 1976; 
Shull 1978; 
Black, Porter \& Dalgarno 1981; 
van Dishoeck \& Black 1986;
Black \& van Dishoeck 1987;
Sternberg 1986, 1988; Sternberg \& Dalgarno 1989).

A number of theoretical investigations have advanced our understanding
of photodissociation regions, or ``PDRs''.\footnote{
	We use ``PDR'' for ``photodissociation region''.  Note, however, 
	that some
	authors use these initials for ``photon-dominated region''.
	}
Solomon (1965: see Field, Somerville, \& Dressler 1966) apparently was the 
first to propose that destruction of interstellar $\HH$ might be dominated
by dissociating transitions to the vibrational continuum following
permitted transitions to the $^1B\Sigma_u^+$ state; the
resulting photodissociation rate was first estimated by 
Stecher \& Williams (1967), who called attention to the potential
importance of $\HH$ self-shielding.
The $\HH$ self-shielding process has been studied by
Shull (1978),
Federman, Glassgold \& Kwan (1979),
and de Jong, Dalgarno, \& Boland (1980).

Tielens \& Hollenbach (1985a) discussed the overall thermal and
chemical structure of photodissociation fronts, and applied their
theoretical model to explain observed properties of the
photodissociation region in Orion (Tielens \& Hollenbach 1985b).
Black \& van Dishoeck (1987) examined in detail the fluorescent
excitation of $\HH$, and the resulting infrared line emission.
Recently Abgrall \etal (1992) presented new data relating to the
photodissociation of $\HH$,
and carried out a more accurate treatment
of the self-shielding process in order to reexamine the structure of
the H/$\HH$ transition zone.
Recent detailed self-shielding calculations
for a cloud of moderate density subject to strong UV illumination
(Le Bourlot \etal 1992) and for diffuse cloud conditions
(Heck \etal 1992) show that in some regions line overlap
significantly affected the $\HH$ photodissociation rate.

In a previous paper (Bertoldi \& Draine 1996)
we discussed the structure
of coupled ionization-dissociation fronts, and concluded that they were
in many cases expected to be propagating at significant velocities,
calling into question the interpretation of some regions, including
Orion, which have been based on models of stationary photodissociation
fronts.
In the course of our investigation we have reexamined the self-shielding
of $\HH$.

In this paper we revisit the problem of $\HH$ self-shielding.
We confirm that line overlap can often be important, and we identify
the region in parameter space where this occurs.
We develop an approximate method to allow for the effects of line overlap
in a statistical fashion.
We apply this method to compute models for photodissociation fronts
including the effects of line overlap; we also improve somewhat upon previous
treatments by applying the results of recent studies of the 
wavelength dependence of dust extinction.
We find a simple analytic description of the $\HH$ self-shielding function
which appears to give a good approximation to the results of detailed
self-shielding calculations, including line overlap.

We construct models of stationary photodissociation fronts for 
different densities, temperatures, and intensities of incident FUV radiation.
We examine how observable properties, such as the 1--0S(1) surface
brightness, and the 1--0S(1)/2--1S(1) and 2--1S(1)/6--4Q(1) line ratios, 
depend on the model parameters.
Two indicators of the excitation mechanism (fluorescent vs. shock)
and the gas density are
the ratio of the ortho-- (odd $J$) to para--H$_2$
(even $J$) level populations and the rotational temperature within a
vibrational level. We compute these quantities as a function of gas
density and FUV illumination.

The famous reflection nebula NGC 2023 is a well-studied example of
fluorescent emission by $\HH$.
We attempt to reconcile published observations of this object.
The available observations of NGC 2023 appear to be consistent with
a photodissociation front with $\nH\approx10^5\cm^{-3}$ irradiated by
FUV radiation with intensity (relative to the Habing field)
$\chi\approx5000$.

\section{ Computation of Level Populations}
\subsection{Equations of Statistical Equilibrium}

Let the $N$ vibration-rotation levels of the ground electronic state of $\HH$
be designated by
the
index $l=1,...,N$.
Then
\begin{eqnarray}
{dn_l \over dt} &=&
R\nH n(\H) \hhini_l +
\sum_{m\neq l}
\left[
A_{lm}
+\beta_{lm}
+\sum_c n_c C_{lm}^{(c)}
\right]
n_m
\nonumber
\\
&~&~~~~~
-n_l\sum_{m\neq l}
\left[A_{ml} + \sum_c n_c C_{ml}^{(c)}\right]
-n_l\left(\beta_{diss,l}+\sum_c n_c C_{diss,l}^{(c)}\right)~,
\label{eq:dnldt}
\end{eqnarray}
\beq
{dn(\H)\over dt} = -2R\nH n(\H) + 2\sum_{m}
\left[ \beta_{diss,m} + \sum_c n_c C_{diss,m}^{(c)}\right]~.
\eeq
The notation is as follows:
$R$ is the rate coefficient for formation of
$\HH$ via grain catalysis;
$\hhini_l$ is the fraction of $\HH$ formed on grain surfaces which
leaves the grain in level $l$;
$A_{lm}$ is the Einstein A coefficient for spontaneous
decay $m\rightarrow l$ ($A_{lm}=0$ for $E_m<E_l$);
$\beta_{lm}$ is the effective rate for transitions $m\rightarrow l$ via
ultraviolet pumping;
$C_{lm}^{(c)}$ is the rate coefficient for transitions $m\rightarrow l$
due to collisions with collision partner $c\in\{\H,\HH,\He,\H^+,e\}$;
$C_{diss,m}^{(c)}$ is the rate coefficient for collisional dissociation out
of level $m$ by collisional partner $c$;
$\beta_{diss,m}$ is the rate of photodissociation out of level $m$.

The effective transition rate from level $m$ to level
$l$ via electronically excited states $u$ due to ultraviolet pumping is
\beq
\beta_{lm}=\sum_u {A_{lu} \over A_{tot,u}}~ \zeta_{um}~~~,
\eeq
where $\zeta_{ul}$ is the rate for photoexcitation to electronically-excited
level $u$ out of level $l$, and
the total rate of spontaneous radiative decay of level $u$ is
\beq
A_{tot,u}=\sum_l A_{lu} + A_{vc,u}~~~;
\eeq
here $A_{vc,u}$ is the rate of spontaneous radiative decay from level $u$
to the vibrational continuum of the ground electronic state, resulting
in dissociation.
The dissociation probability for level $u$ is
\beq
p_{diss,u}=A_{vc,u}/A_{tot,u}~,
\eeq
and the rate for photodissociation from level $m$ is
\beq
\beta_{diss,m}=\sum_u p_{diss,u}\zeta_{um}~.
\eeq
The dissociation probability averaged over photoexcitations out
of level $m=X(v,J)$ is
\beq
\langle p_{diss}(v,J)\rangle=\beta_{diss,m}/\sum_u\zeta_{um}~~~.
\eeq
It is convenient to define the diagonal elements $A_{ll}$, $C_{ll}$,
and $\beta_{ll}$ to be
\beq
A_{ll} \equiv - \sum_{m\neq l}A_{ml}~,
\eeq
\beq
\beta_{ll} \equiv - \sum_{m\neq l}\beta_{ml} - \beta_{diss,l}~,
\eeq
\beq
C_{ll}^{(c)} \equiv - \sum_{m\neq l}C_{ml}^{(c)} - C_{diss,m}^{(c)}~.
\eeq
Eq. (\ref{eq:dnldt}) then becomes
\beq
{dn_l\over dt} = R\nH n(\H)\hhini_l + \sum_m D_{lm}n_m~,
\eeq
\beq
{dn(\H)\over dt} = -2R\nH n(\H) + 2\sum_m D_{diss,m}n_m~,
\eeq
where
\beq
D_{lm} \equiv A_{lm} + \beta_{lm} + \sum_c n_c C_{lm}^{(c)}~,
\eeq
\beq
D_{diss,m} \equiv \beta_{diss,m} + \sum_c n_c C_{diss,m}^{(c)}~.
\eeq
We include the 299 bound states of $\HH$ with $J\leq29$ in our calculations.

\subsection{Radiative Rates \label{sec:radrates}}

We designate the vibration-rotation levels of the
ground electronic state $X^1\Sigma_g^+$ by $X(v,J)$, and the
first 3 electronically-excited states $B^1\Sigma_u^+$, $C^1\Pi_u^+$ and
$C^1\Pi_u^-$ by $B(v,J)$, $C^+(v,J)$, and $C^-(v,J)$.

Energy levels $E_u$, transition probabilities $A_{lu}$, and dissociation 
probabilities $p_{diss,u}$ have been published by Abgrall \& Roueff (1989) and
Abgrall \etal (1992,1993a,b).  
We used data generously provided by Roueff (1992), covering levels up
to $J=29$.
As discussed by Abgrall \etal (1992), these new data include dissociation
probabilities $p_{diss,u}$ 
for the $C^+(v,J)$ levels which are considerably larger
than the dissociation probabilities of Stephens \& Dalgarno (1972), which
were used for both $C^+(v,J)$ and $C^-(v,J)$ in earlier detailed modelling
(e.g., van Dishoeck \& Black 1986; Black \& van Dishoeck 1987).
As a result, we find somewhat larger overall dissociation probabilities.

The transition probabilities $A_{lu}$ and dissociation
probabilities $p_{diss,u}$ are assumed to be independent of the energy of
the photon responsible for photoexcitation to level $u$: the branching
ratios for radiative decay out of level $u$ are assumed to {\it not} depend on
whether the photoexcitation to level $u$ was due to a photon which was
precisely resonant ($\lambda=\lambda_{lu}$) or far out on the damping
wings of the transition (e.g., $\lambda=1.001\lambda_{lu}$).
While this approximation is usual, it should be remembered that it
is probably not exact.

\subsection{Collisional Rates \label{ss:collrates}}
We include inelastic collisions with H, He, $\HH$, H$^+$, and $e^-$, with
collisional rate coefficients as described in Draine \& Bertoldi (1996).
For H--$\HH$ collisions,
inelastic cross sections have been computed by Mandy \& Martin (1993,1996)
using semiclassical trajectory calculations; Martin \& Mandy (1995)
have provided fits to the resulting rate coefficients
for $450<T<20000\K$.
In the present application it is necessary to extrapolate to
temperatures $T<450\K$.
The Martin \& Mandy fitting function is
\beq
\log_{10}\langle\sigma v\rangle = a + bz
+ cz^2 - d\left({4500\K\over T}-1\right)~~~,
\label{eq:mmfit}
\eeq
where $z\equiv\log_{10}(T/4500\K)$.
Even for downward transitions, the coefficient $d$ is typically large enough
that this function declines very rapidly with decreasing temperature
at low temperatures.
Furthermore, comparison of semiclassical trajectory calculations and
quantal calculations indicate that the former approximation
substantially underestimates the inelastic cross sections near the
energy threshold (Lepp, Buch \& Dalgarno 1995).
Accordingly, we use eq.\ (\ref{eq:mmfit}) for $T>\theta=600\K$,
but for $T<\theta$ we take
\beq
\log_{10}\langle\sigma v\rangle = a+bz
+cz^2 -d\left({9000\K\over\theta} - 1 - {4500\K T\over\theta^2}\right)~~~.
\label{eq:mmextrap}
\eeq
This expression joins smoothly to eq.\ (\ref{eq:mmfit}) at $T=\theta$,
but declines less dramatically for $T<\theta$.\footnote{
	Even so, the resulting rate coefficients for $(0,3)\rightarrow(0,1)$ 
	and (0,2)$\rightarrow$(0,0)
	at $T=200\K$ are only $2.9\times10^{-14}\cm^3\s^{-1}$ and
	$1.4\times10^{-14}\cm^3\s^{-1}$, smaller than the
	best estimates of Lepp et al. by factors of 30 and 100, respectively.
	}
Since we have extrapolated in a rather arbitrary manner, it is clear that
quantal calculations  with an accurate potential surface are needed to
provide better estimates of the low temperature H-$\HH$ rate coefficients.

H$^+$ is treated as a species with a fixed abundance
$x_\H\equiv n(\H^+)/\nH=1\times10^{-4}$;
we assume\footnote{
	For a photoionized metal abundance $n(M^+)/\nH=2\times10^{-4}$
	(primarily C$^+$), 
	a hydrogen ionization rate
	$\zeta_{\rm H}=
	8\times10^{-18}(T/10^2\K)^{-0.64}(\nH/40\cm^{-3})\s^{-1}$
	(due to either cosmic rays or X-rays) would sustain a
	hydrogen ionization fraction $x_\H=1\times10^{-4}$.}
$n_e/\nH=3\times10^{-4}$.

\subsection{H$_2$ Formation on Grains\label{sec:h2form}}

As discussed elsewhere (Draine \& Bertoldi 1996),
we assume that
H atoms recombine on grain surfaces to
form $\HH$ at
a rate per volume
\beq
\left({dn(\HH)\over dt}\right)_{\rm form} = R~\nH n(\H)~~~,
\eeq
with
\beq
R = 6\times10^{-18}~T^{1/2}\cm^3\s^{-1}~~~;
\label{eq:h2form}
\eeq
this adopted value of $R$ is $2/3$ of the value adopted by
Black and van Dishoeck (1987) for most of their model calculations.
For $T=70\K$ this rate is $\sim1.7$ times larger than the value of
$R$ inferred by Jura (1975) from {\it Copernicus} observations of $\HH$ in
diffuse clouds.

The
rovibrational distribution
$\hhini(v,J)$ of the newly-formed $\HH$ is taken to be of the form
(Draine \& Bertoldi 1996)
\beq
\hhini(v,J) = \hhini(0,0)g_n(J)(1+v)\exp[-E(v,J)/kT_{\rm f}]~,
\label{eq:phi}
\eeq
where $g_n=1,3$ for even,odd $J$.
We note that this distribution (\ref{eq:phi})
does {\it not} include a rotational
degeneracy factor $(2J+1)$ and {\it does} include a factor $(1+v)$; these
deviations from a thermal distribution function are intended to enhance
the populations of high-$v$ states relative to high-$J$ states.
In the present paper we take 
$T_{\rm f}=5\times10^4\K$, in which case the newly-formed $\HH$ has
an ortho/para ratio = 2.78,
$\langle v\rangle=5.3$,
$\langle J\rangle=8.7$,
$\langle J^2\rangle^{1/2}=10.7$,
and a mean vibration-rotation energy of $2.89\eV$.

\section{Radiation Field \label{ss:radfld}}

The efficacy of $\HH$ self-shielding depends upon the distribution of
$\HH$ over the different levels $X(v,J)$, which in turn depends upon the
density and temperature of the gas, and the incident radiation field.

In neutral regions, pumping of $\HH$ is primarily effected by
far-ultraviolet photons in the $1110-912\Angstrom$ range.\footnote{
	\label{footnote:1110-912}
	Photons shortward of $912\Angstrom$ are of course absorbed by atomic 
	hydrogen.
	$\HH$ in the $v=0$ levels has only weak absorptions longward of
	$1110\Angstrom$: the longest wavelength absorption out of the
	$J=0$ and 1 levels is the Lyman 0-0P(1) line at $1110.066\Angstrom$,
	with an oscillator strength $f_{lu}=0.00058$
	(as compared to the much stronger 7-0R(1)
	(i.e. $v_u=7$, $v_l=0$, $J_u-J_l=1$, $J_l=1$)
	line at $1013.434\Angstrom$,
	with $f_{lu}=0.020$).
	While there are weak absorptions out of $v=0,J=2,3,4,5,...$ longward
	of $1110\Angstrom$ (Lyman 0-0P(2),P(3),P(4),P(5),... at $1112.5$,
	$1115.9$, $1120.3$, $1125.5\Angstrom$...,
	with $f_{lu}=0.00070$, $0.00074$, $0.00076$, $0.00076$...,
	it is fair to assume
	that the bulk of UV pumping of $\HH$ is due to photons in the
	1110-912$\Angstrom$ interval.
	}
Habing (1968) estimated the intensity of interstellar starlight at
$\lambda=1000\Angstrom$ to be $\lambda u_\lambda=4\times10^{-14}\erg\cm^{-3}$.
We will consider various radiation fields $u_\lambda$, and will 
characterize the intensity of each at $1000\Angstrom$, relative to
Habing's estimate, by the dimensionless 
factor
\beq
\chi \equiv {[\lambda u_\lambda]_{1000\Angstrom} \over
4\times10^{-14}\erg\cm^{-3}}~~~.
\label{eq:chidef}
\eeq
In Table 1 we list spectral forms which
have been considered by various workers.
For each spectrum we list the photon flux
in the $1110-912\Angstrom$ interval,
\beq
F \equiv \int\limits_{912\Angstrom}^{1110\Angstrom} h^{-1}
\lambda u_\lambda d\lambda~~~,
\label{eq:fdef}
\eeq
the logarithmic derivative $d\ln u_\nu/d\ln\nu$ evaluated at $1000\Angstrom$,
and the color temperature $T_{color}$, the temperature of a blackbody having
the same logarithmic derivative at $1000\Angstrom$.

Habing (1968) estimated the average radiation field at 
1000, 1400, and 2200\AA.
We note that Habing's values can be fitted by
\beq
\lambda u_\lambda=(-25\lambda_3^3/6+25\lambda_3^2/2-13\lambda_3/3)
\times10^{-14}\erg\cm^{-3}~~~,
\label{eq:habing}
\eeq
where $\lambda_3\equiv\lambda/10^3\Angstrom$;
this is the spectrum listed as ``Habing'' in Table 1.

Draine (1978) approximated UV starlight by the
relatively soft radiation field
\beq
\lambda u_\lambda = 4\times10^{-14}\chi \lambda_3^{-5}
\left[ 31.016\lambda_3^2-49.913\lambda_3
+19.897\right]
\erg\cm^{-3}~~~,
\label{eq:draine78}
\eeq
with $\chi=1.71$.

Roberge, Dalgarno, \& Flannery (1981) used a radiation field similar to
(\ref{eq:draine78}).
This radiation field was employed by van Dishoeck \& Black (1986) in their
modelling of diffuse clouds, and by
Black \& van Dishoeck (1987) in their models of photodissociation fronts.
As seen in Table 1, this spectrum is somewhat harder than 
eq.\ (\ref{eq:draine78}).

As an example of the radiation field expected in a photodissociation
front near an OB star, we consider the
power-law spectrum $u_\nu\propto\nu^{-2}$:
\beq
\nu u_\nu (x=0) \equiv
\lambda u_\lambda (x=0) = 4\times 10^{-14}\chi \lambda_3
\erg\cm^{-3}~~~.
\label{eq:u_lambda}
\eeq
This spectrum has $T_{color}=29000\K$, corresponding to the spectrum of
a B0 star.
We use this spectrum in most of the model calculations; as we show
below (\S\ref{sec:incident_spectrum}, and 
Fig.~\ref{fig:tcolor_and_tform}), the properties of the PDR are 
insensitive to the spectrum for color temperatures
$T_{color}\gtsim10^4\K$, so that the models presented here have
general applicability to PDRs illuminated by stars of spectral type
A0 and 
earlier.\footnote{
	Late B-type stars of course do not have large luminosities in the
	1110--912\AA\
	interval, and can only produce large $\chi$ values
	close to the star.
	}

Pumping rates and dissociation rates for optically-thin $\HH$ in
various rotation-vibration states are given in Table 2 for the
radiation fields (\ref{eq:u_lambda}) and (\ref{eq:draine78}).
Also given in Table 1 is the probability $p_{diss}(v,J)$ that photoexcitation
out of level $X(v,J)$ will be followed by dissociation, and
the probability $\langle p_{ret}(v,J)\rangle$ 
that photoexcitation out of $X(v,J)$ will
be followed by direct spontaneous decay back to the original level
$X(v,J)$.
Since there are many possible UV transitions possible out of any given
level $X(v,J)$, both $p_{diss}$ and $\langle p_{ret}\rangle$ 
depend on the shape of the
illuminating spectrum, but Table 2 shows that for unshielded $\HH$
the changes in
$\langle p_{diss}\rangle$ or $p_{ret}$ are typically only a few percent when the
shape of the radiation field is changed.
Note, however, that changes in the radiation field due to self-shielding
can lead to larger changes in, for example, $\langle p_{diss}\rangle$, as the
relative importance of inherently
weak absorption transitions increases.
Indeed, for complete photodissociation fronts we find 
(see \S\ref{sec:results}) that 
$\langle p_{diss}\rangle\approx0.15$ averaged over all UV absorptions.

Shull (1978) has noted that when the UV intensity is high, photoexcitation
out of vibrationally-excited levels can affect the ``infrared cascade''
following UV pumping.
For example, with no shielding the rate of photoexcitation
out of the $X(1,3)$ level is $3.30\times10^{-10}\chi\s^{-1}$
for the spectrum (\ref{eq:u_lambda}).
Since the total rate for spontaneous decay out of $X(1,3)$ is
$8.35\times10^{-7}\s^{-1}$, we see that 
in the optically-thin part of the photodissociation front 
($N(\HH)\ltsim10^{14}\cm^{-2}$)
UV pumping will significantly
interfere with the ``infrared cascade'' for $\chi\gtsim2500$.
Our calculations explicitly include UV pumping out of all levels
$X(v,J)$.

\section{Shielding of H$_2$}

\subsection{Shielding by Dust\label{sec:dust_shield}}

The wavelength-dependence of interstellar extinction has been
discussed by Cardelli, Clayton \& Mathis (1989, hereafter CCM) 
and O'Donnell (1994), and found to be very nearly a one-parameter family,
where the parameter may be taken to be $R_V\equiv A_V/E(B-V)$.
The extinction at $\lambda>9000\Angstrom$ appears to be ``universal'',
with an extinction cross section per H nucleus 
$\sigma_{ext}(9000\Angstrom)=2.4\times10^{-22}\cm^2$.
The ``average'' extinction curve has $\NH/E(B-V)=5.8\times10^{21}\cm^{-2}$
(Bohlin, Savage \& Drake 1978) and $R_V\equiv A_V/E(B-V)=3.1$; 
the CCM extinction curve
then predicts an extinction cross section per H of
$\sigma_{ext}(1000\Angstrom)=2.62\times10^{-21}\cm^2$.
In dense molecular regions, however, the extinction appears to typically have
larger values of $R_V = 4-6$ (Mathis 1990), 
with a less steep rise in the ultraviolet.
We will consider two cases: $R_V=3.1$ (for diffuse clouds)
and $R_V=5.5$ (the value observed toward $\theta^1$Ori C
in the Trapezium, possibly representative of very dense clouds);
for $R_V=5.5$ CCM predict
$\sigma_{ext}(1000\Angstrom)=8.16\times10^{-22}\cm^2$.

We consider plane-parallel clouds, with the radiation incident
normally on {\it one} cloud surface.
We do not attempt to treat scattering in any detail, but simply
assume that the effect of dust is to attenuate the incident radiation
field by a factor
$\exp(-\tau_{d,\lambda})$, where
$\tau_{d,\lambda}=\NH \sigma_{d,\lambda}$;
$\NH=N(\H)+2N(\HH)+N(\H^+)$ is the total hydrogen column density
between the point of interest and the cloud surface,
and $\sigma_{d,\lambda}$ is the effective attenuation cross section at
wavelength $\lambda$.

The dust albedo in the $1110-912\Angstrom$ region is uncertain; a recent
study of the reflection nebula NGC7023 near $\lambda=1000\Angstrom$ finds
an albedo $\sim 0.4$, and a scattering asymmetry factor
$\langle\cos\theta\rangle\approx0.75$
(Witt \etal 1993).
The effects of scattering are of course complicated and geometry-dependent,
but if the albedo is 0.4 then
$0.6<\sigma_d/\sigma_{ext}<1$.
At $\lambda=1000\Angstrom$ we take
the effective attenuation cross section to be
$\sigma_{d,1000}=2\times10^{-21}\cm^2$ for $R_V=3.1$
and $6\times10^{-22}\cm^2$ for $R_V=5.5$ -- about 75\% of the estimated
extinction cross sections at this wavelength.
At wavelengths other than $1000\Angstrom$ we take the attenuation
cross section $\sigma_{d,\lambda}=(A_\lambda/A_{1000})\sigma_{d,1000}$,
where the ratio of extinction cross sections $A_\lambda/A_{1000}$ is
given by the CCM extinction curve for $R_V=3.1$ or $R_V=5.5$.

\subsection{Self-Shielding Neglecting Line Overlap\label{sec:nooverlap}}

Each level $X(v,J)$ has many allowed transitions to vibration-rotation
levels of the B, C$^+$, and C$^-$ electronic states.
Let subscripts $l$ and $u$ denote the lower and upper levels.
Let a continuum radiation field be
incident on the cloud, with specific
energy density $u_\lambda$ at the cloud surface.
Neglecting line overlap, the
rate of photoexcitation $l\rightarrow u$ for an $\HH$ molecule in level $l$ is
simply
\beq
\zeta_{ul}(N_l)= {\lambda_{ul}^2 u_\lambda  \over h}
\left({d W_{ul}\over dN_l}\right)
\exp\left[-\tau_d(\nu_{ul})\right]~,
\label{eq:nooverlap}
\eeq
where $N_l$ is the column density of $\HH$ in level $l$,
$\tau_d(\nu)$ is the ``attenuation'' optical depth due to dust,
and $W_{ul}(N_l)$ is the dimensionless equivalent width for the 
transition $l\rightarrow u$:
\beq
W_{ul}(N_l) = \int 
{d\lambda\over\lambda} 
\left[ 1-\exp(-N_l\sigma_{ul}) \right]~,
\label{eq:eqwdth}
\eeq
where the photoabsorption cross section 
$\sigma_{ul}(\lambda)$,
given by the Voigt line profile function,
depends upon the oscillator strength $f_{lu}$ for the transition,
the usual Doppler broadening parameter 
$b={\rm FWHM}/\surd 4\ln 2$,
and the intrinsic broadening $A_{tot,u}$ of the upper level.
We use accurate numerical
approximations for $dW_{ul}/dN_l$ and
$W_{ul}(N_l)$ given by Rogers \& Williams (1974).

\subsection{Self-Shielding Including Line Overlap\label{sec:withoverlap}}

For small column densities, equation (\ref{eq:nooverlap})
will give a good estimate of the effects of self-shielding.
At large column densities, however, lines may be shielded in part by
overlap with other lines, as noted by Black \& Dalgarno (1977).
Many previous treatments have neglected this overlap: Black \& van Dishoeck
(1987) cited Roberge (1981) and Sternberg (1986) as having shown line
overlap to be negligible for normal dust-to-gas ratios.
More recently, detailed transfer calculations (Abgrall \etal 1992;
Le Bourlot \etal 1992) have found that line overlap can become
important under some conditions.
Here we reconsider the importance of line overlap.

Exact treatment of the effects of line overlap
requires explicit calculation of the
full radiation field as a function of frequency at each point.
However, a simple estimate for the effects of line overlap may be obtained
by first noting that the strong $\HH$ absorption lines originate from
the $v=0$ level and will therefore be almost entirely confined to
the wavelength range $1110-912\Angstrom$
(see footnote \ref{footnote:1110-912}).
Therefore for lines with $\lambda_{ul}>1110\Angstrom$ we will neglect
line overlap, and use equation (\ref{eq:nooverlap}).

We note that the wavelength range $1110-912\Angstrom$ includes
H Ly$\beta,\gamma,\delta,...$ at 1025.72, 972.54, 949.75, ...$\Angstrom$.
Since the PDR may develop large column densities $N(\H)$, these lines
may suppress the pumping rates for nearby $\HH$ lines.

For the lines with $912<\lambda_{ul}<1110\Angstrom$, we note that
the total equivalent width of the ensemble of lines, including H lines and
the effects of line
overlap, will asymptotically approach $W_{max}\approx\ln(1110/912)\approx0.2$
at large column densities.
If we now treat the lines ``democratically'' -- equally subject to
the effects of line overlap -- then we
may approximate the effects of line overlap by
considering a modified equivalent width $\tilde{W}_{ul}$
defined by
\beq
{d\tilde{W}_{ul}\over dN_l} = {dW_{ul}\over dN_l}
\exp(-W/W_{max})~~~{\rm for}~912<\lambda_{ul}<1110\Angstrom~,
\eeq
where
\beq
W\equiv \sum_l{\sum_u}^\prime W_{ul} + 
\sum_{n=3}^{15} 
W(\H 1-n)~~~{\rm for}~912<\lambda_{ul}<1110\Angstrom~;
\label{eq:w}
\eeq
the prime indicates that we restrict the sum to transitions with
$1110 < \lambda_{ul} < 912\Angstrom$.
For $n\gtsim10$ the H Lyman damping wings are not important, and the
Lyman series lines blend together to form a continuum for
$n\gtsim 39(3\kms/b)^{1/3}$, or 
$\lambda\ltsim (911.76+0.60(b/3\kms)^{2/3})\Angstrom$.
We will somewhat arbitrarily consider the H Lyman lines individually
up to $n=15$; 
the $\H1-15$ line is at $915.83\Angstrom$.
The damping constants $A_{tot}$ for the $np$ levels were obtained by
summing over $A$ coefficients for the different decay channels; the
latter were computed using radial integrals tabulated by
Green, Rush, \& Chandler (1957).
We will see below that absorption by the H Lyman lines has only a small
effect on the $\HH$ pumping rates.

It is easily seen that
\beq
\tilde{W}\equiv
\sum_u{\sum_l}^\prime\tilde{W}_{ul} 
+ \sum_{n=3}^{15}\tilde{W}(\H1-n)
= W_{max}\left[1-\exp(-W/W_{max})\right]~~~,
\eeq
so that $\tilde{W}\rightarrow W_{max}$ as $W\rightarrow\infty$.
To allow for line overlap, therefore, we take the pumping rates to be
\beq
\zeta_{ul}(N_l)={\lambda_{ul}^2 u_\lambda  \over h}
\left({d W_{ul}\over dN_l}\right)\exp(-W/W_{max})
\exp\left[-\tau_d(\lambda_{ul})\right]~~~,
\label{eq:withoverlap}
\eeq
with 
$W_{ul}$ given by eq.\ (\ref{eq:eqwdth}, $W$ given by eq.\ (\ref{eq:w}), and 
$W_{max}=0.2$.

The pumping rate and photodissociation rate out of level $l$ are simply
\beq
\zeta_{pump,l}=\sum_u \zeta_{ul}~~~,
\eeq
\beq
\zeta_{diss,l}=\sum_u \zeta_{ul}p_{diss,u}~~~.
\eeq
The effective pumping rate and photodissociation rate per $\HH$ molecule 
are just
\beq
\zeta_{pump}=[n(\HH)]^{-1}\sum_l \zeta_{pump,l}n_l~~~,
\eeq
\beq
\zeta_{diss}=[n(\HH)]^{-1}\sum_l \zeta_{diss,l}n_l~~~,
\eeq
where $n_l$ is the number density of $\HH$ in level $l$.

\subsection{Approximations for Self-Shielding}

For many purposes it is useful to have simple analytic approximations
for the dependence of the pumping rate on $N_2\equiv N(\HH)$ and
dust extinction $\tau_{d,1000}=\NH\sigma_{d,1000}$.
Since the dust extinction does not vary strongly over the $1110-912\Angstrom$
range, the pumping rate is approximately
\beq
\zeta_{pump}(N_2)\approx f_{shield}(N_2)e^{-\tau_{d,1000}}\zeta_{pump}(0)
\eeq
where $f_{shield}(N_2)$ represents the self-shielding effect of the
$\HH$.
Here we put forward two approximations for $f_{shield}(N_2)$; 
their accuracy will be assessed later by comparison with detailed calculations.

A very simple power-law approximation is provided by
\beq
f_{shield}(N_2)=
\cases{
	1 & for $N_2 < 10^{14}\cm^{-2}$, \cr
	(N_2/10^{14}\cm^{-2})^{-0.75} & 
		for $N_2 > 10^{14}\cm^{-2}$.\cr
	}
\label{eq:powfit}
\eeq
The $N_2^{-0.75}$ dependence is steeper than the $N_2^{-0.5}$ behavior
expected for heavily saturated lines, but we shall see 
below in \S\ref{sec:results} that
it approximates $f_{shield}$ quite well over a large range of column
densities.
Note, however, that this approximation behaves unphysically
in the limit $N_2\rightarrow\infty$: if there were no dust, the total
pumping
rate per area $\int_0^{\infty}\zeta_{pump}dN_2 \rightarrow\infty$.
This unphysical behavior corresponds to a failure to require that
the total equivalent width be limited to some value $W_{max}\approx0.2$.
We can correct this deficiency with a more complicated expression:
\beq
f_{shield}(N_2) =
{0.965\over (1+x/b_5)^2}
+
{0.035\over (1+x)^{0.5}}
\exp\left[-8.5\times10^{-4}(1+x)^{0.5}\right]~,
\label{eq:goodfit}
\eeq
where $x\equiv N_2/5\times10^{14}\cm^{-2}$,
and $b_5\equiv b/10^5\cm\s^{-1}$.
This somewhat complicated functional form has been constructed so that
in the absence of dust the total pumping rate per area
$\int_0^{\infty}\zeta_{pump} dN_2 \approx\chi F$
for the $u_\nu\propto \nu^{-2}$ spectrum (\ref{eq:u_lambda}).
It is also designed to reproduce the complete transition from optically-thin
to extreme saturation with line overlap.
We shall see below that equation (\ref{eq:goodfit}) 
is significantly more accurate than equation (\ref{eq:powfit}), yet is
nevertheless straightforward to evaluate numerically and hence 
useful for numerical modelling.
\begin{figure}[t]
\epsscale{0.60}%size1
\plotone{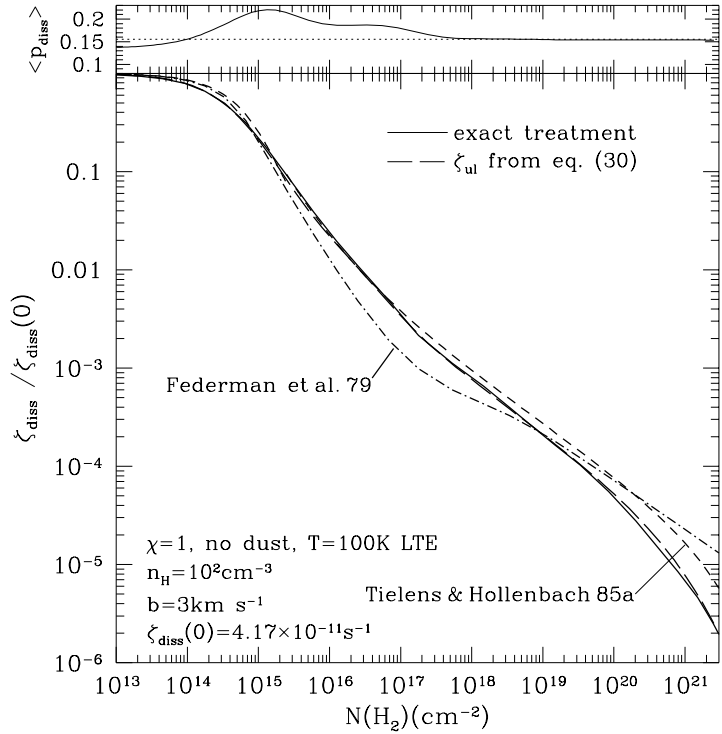}
\caption{
	Comparison of self-shielding calculated using the statistical
	treatment of line overlap 
	(eq.\protect\ref{eq:withoverlap})
	and an ``exact'' treatment of line overlap,
	for $T=100\K$ LTE level populations.
	The $\HH$ is assumed to have a Doppler broadening parameter
	$b=3\kms$.
	Models do not include dust shielding or absorption by CI.
	It is apparent that 
	eq. (\protect\ref{eq:withoverlap})
	does an excellent job of accounting for
	line overlap.
	Also shown are self-shielding approximations of Federman \etal (1979)
	and Tielens \& Hollenbach (1985a) (see text).
	The upper panel shows the dissociation probability 
	$\langle p_{diss}\rangle$ as a function of position; the dotted line
	shows the dissociation probability averaged over all UV pumping
	events in the PDR.
	\label{fig:LTE100}
	}
\end{figure}
\begin{figure}[t]
\epsscale{0.60}%size1
\plotone{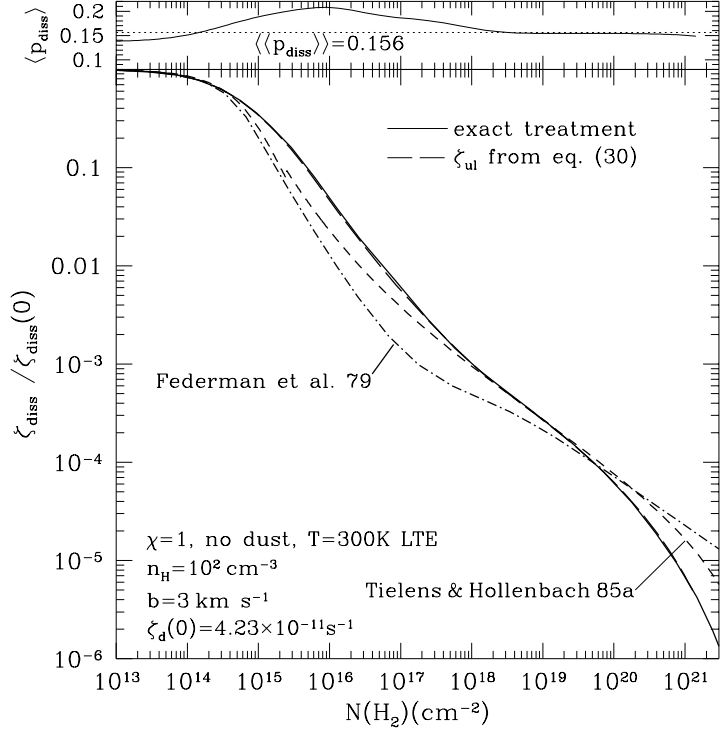}
\caption{
	Same as Figure 1, but for $T=300\K$ LTE populations..
	\label{fig:LTE300}
	}
\end{figure}
\begin{figure}[t]
\epsscale{0.60}%size1
\plotone{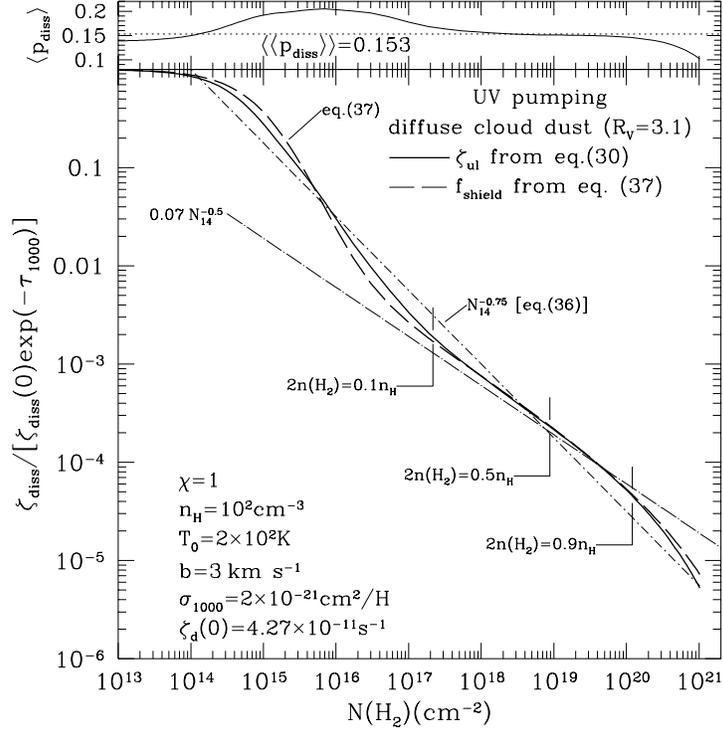}
\caption{
	$\HH$ self-shielding factor $f_{shield}$
	as a function of $N(\HH)$, in a
	stationary photodissociation front with $\chi/\nH=.01\cm^3$
	and $T_0=200\K$.
	The $\HH$ is assumed to have a Doppler broadening parameter
	$b=3\kms$.
	Dust absorption is included assuming diffuse cloud dust with
	$R_V=3.1$.
	The solid line shows the dissociation rate computed from a multiline
	calculation with approximate treatment of line overlap, using
	eq.\ (\protect\ref{eq:withoverlap}).
	The long broken line is the semi-analytic fit 
	(\protect\ref{eq:goodfit}).
	Two power-law fits are shown: $N_{14}^{-0.75}$ 
	(eq.\protect\ref{eq:powfit}),
	and a square-root fit (see text).
	The upper panel shows the dissociation probability
	$\langle p_{diss}\rangle$; the dotted line shows the dissociation
	probability averaged over all UV pumping events in the PDR.
	\label{fig:chi/nh=.01,T_0=200,Rv=3.1}
	}
\end{figure}
\begin{figure}[t]
\epsscale{0.60}%size1
\plotone{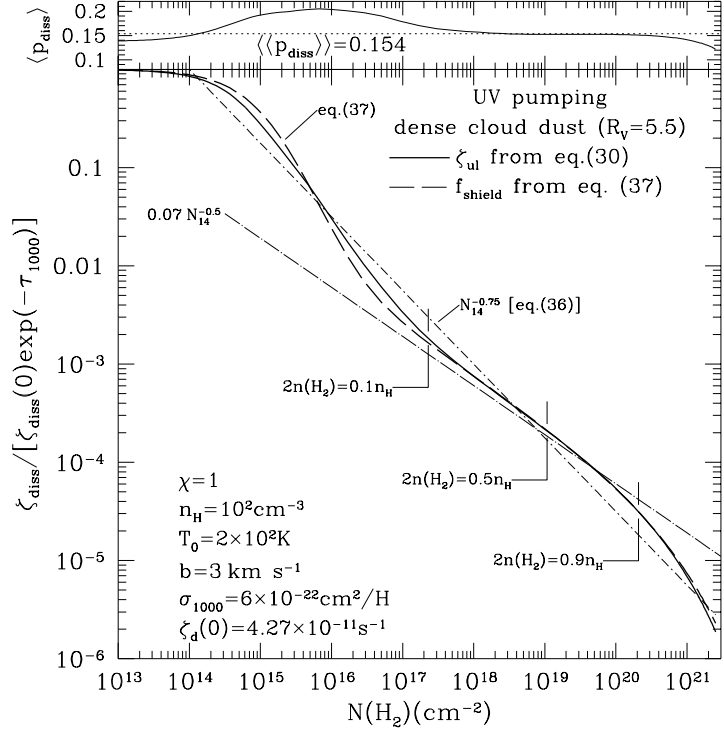}
\caption{
	Same as Fig.\ 3, but for dense cloud dust with $R_V=5.5$.
	\label{fig:chi/nh=.01,T_0=200,Rv=5.5}
	}
\end{figure}
\begin{figure}[t]
\epsscale{0.60}%size1
\plotone{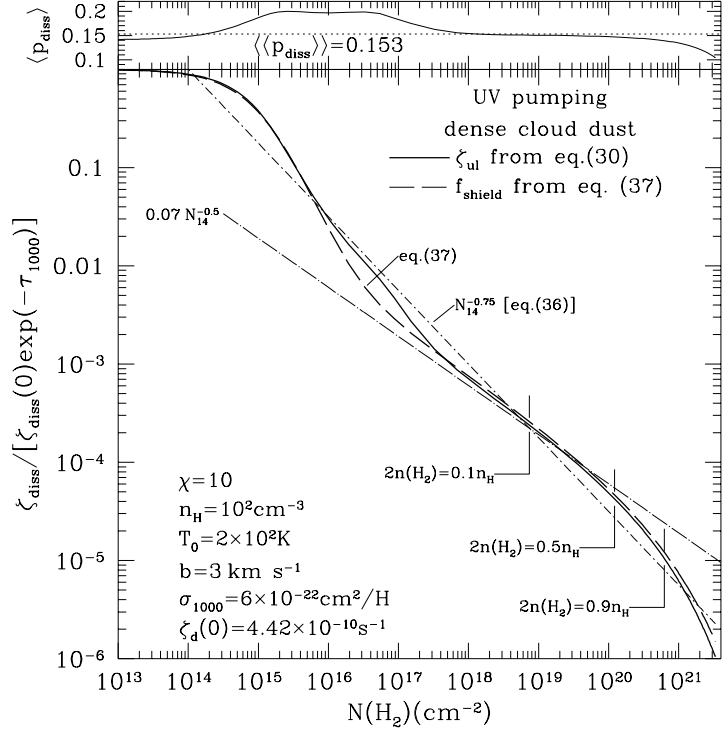}
\caption{
	Same as Fig.\ 3, but for dense cloud dust and a
	stronger radiation field, with $\chi/\nH=0.1\cm^3$.
	\label{fig:chi/nh=.1,T_0=200,Rv=5.5}
	}
\end{figure}
\begin{figure}[t]
\epsscale{0.60}%size1
\plotone{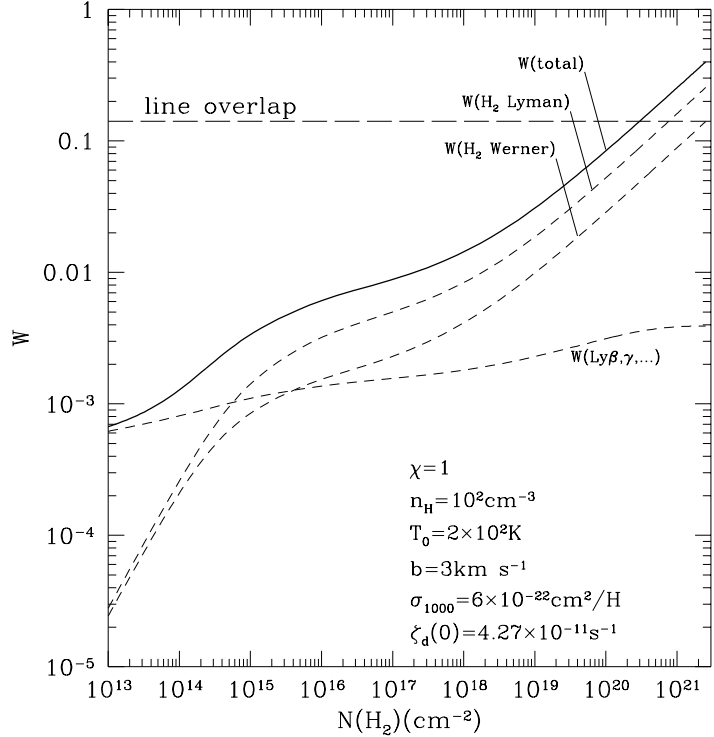}
\caption{
	Total dimensionless equivalent width $W$ contributed
	by lines in the $1110-912\Angstrom$ interval, as a function of the
	$\HH$ column density, for $\chi/\nH=0.01\cm^3$, $b=3\kms$, and
	dense cloud dust.
	Separate contributions of Lyman band, Werner band, and
	HI Lyman series (Ly$\beta$ through Lyman-15) are shown.
	Line overlap is expected to suppress pumping rates by a factor $\sim2$
	when $W(total)$ equals $0.2\ln(2)=0.14$; this occurs at
	$N(\HH)=3\times10^{20}\cm^{-2}$.
	\label{fig:eqwidths}
	}
\end{figure}

Since the pumping rate may be written 
$\zeta_{pump}=F_\nu d W_\nu/dN_2 = \nu F_\nu dW/dN_2$,
(where $F_\nu=c u_\nu / h \nu$)
and $\zeta_{pump}=\zeta_{pump}(0)~ f_{shield}$, we obtain
\beq
W(N_2) =
\Delta\ln\nu~~{\zeta_{pump}(0)\over F} 
\int_0^{N_2} dN_2^\prime f_{shield}(N_2^\prime)
\eeq
\beq
=
\Delta\ln\nu ~~~0.947 \left\{ 1+ {.0117x\over 1+x/b_5} -
\exp\left[-8.5\times10^{-4}\left((1+x)^{0.5}-1\right)\right]\right\}~,
\eeq
thereby demonstrating that our assumed shielding function (\ref{eq:goodfit})
corresponds to an
equivalent width $W\approx \Delta\ln\nu \approx 0.2$ in the limit
$N_2\rightarrow\infty$.
Neglecting variations in the dissociation 
probability $\langle p_{diss}\rangle$, we may approximate the 
photodissociation rate by
\beq
\zeta_{diss}\approx f_{shield}(N_2)e^{-\tau_{d,1000}}\zeta_{diss}(0) ~~~.
\eeq

\section{Structure of Stationary Photodissociation Fronts}
 
\subsection{Model Assumptions}
 
We now consider models of stationary photodissociation fronts, in which
the $\HH$ level populations at each point are the steady-state
level populations resulting from
UV pumping, spontaneous radiative decay, collisional
excitation and deexcitation,
photodissociation, and $\HH$ formation on grains.

In the models presented here we do not examine the thermal balance in 
the PDR.
Theoretical estimates for the temperature structure of PDRs
are very uncertain, primarily due to uncertainties in the grain
photoelectric heating: compare the temperature profiles of 
Tielens \& Hollenbach (1987), Burton, Hollenbach \& Tielens (1990), and
Bakes \& Tielens (1994).
We will simply assume 
the kinetic temperature in the PDR to vary as
\beq
T = {T_0\over 1 + \tau_{d,1000}} ~~,
\label{eq:tprof}
\eeq
where the initial temperature $T_0$ will be treated as an adjustable
parameter.
This temperature profile has the expected property of declining at large
optical depth, as the dust grains attenuate the incident UV.
The functional form is not intended to be realistic, but does allow us
to explore the effects of varying the gas temperature in the region where
$\HH$ pumping takes place.

We assume a plane-parallel geometry, with radiation propagating in the
$+x$ direction.
The UV intensity at $x=0$ is specified by the parameter $\chi$.
At each point we assume an equilibrium between $\HH$ formation on grains
and $\HH$ photodissociation.
The $\HH$ level populations are similarly assumed to be in steady-state
statistical equilibrium.
If we define
\beq
y_l \equiv 2n_l/\nH
\eeq
and require $\dot{y}_l=0$,
then, since $n(\H)=\nH(1-x_\H-\sum_{l=1}^N y_l)$, the vector $y_l$ must
satisfy the system of $N$ inhomogeneous equations
\beq
\sum_{m=1}^N (2R\nH\hhini_l - D_{lm})y_m = 2R\nH(1-x_\H)\hhini_l ~~~.
\eeq
As discussed in \S2, we 
set $x_\H=1\times10^{-4}$, and
include the $N=299$ bound states of $\HH$ with
$J\leq29$ in our calculations.
We solve for the level populations $y_m, m=1,...,N$
by LU decomposition,  using routines LUDCMP and LUBKSB from Press et al.
(1992).
We will refer to these models as ``UV-pumping'' models, since the 
rovibrational distribution of $\HH$ is strongly affected by the
UV pumping, and, as a result: 
(1) there is a significant population in
rotational levels $J > 5$ which would be negligibly populated in
a LTE distribution with
$T_{exc}\approx100\K$;
(2) the rovibrational distribution of the $\HH$ varies with depth
in the cloud, since pumping effects are strongest near the cloud surface.
 
For comparison, we will also consider models in which there is a
balance between $\HH$ photodissociation and formation on grains, but
where the distribution of the absorbing $\HH$ over rotation-vibration levels
is assumed to be a thermal distribution with a specified
excitation temperature $T_{exc}=100\K$.
These will be referred to as ``LTE'' models.
 
\subsection{Self-Shielding\label{sec:results}}
 
In order to test the accuracy of our statistical treatment of
line overlap, we have carried out detailed (``exact'') radiative
transfer calculations to compare with the photoexcitation rates
computed using our statistical treatment of line overlap 
[eq.\ (\ref{eq:withoverlap})].  The ``exact'' calculations explicitly
took into account the 900 strongest FUV absorption lines of $\HH$,
with a fine enough
frequency sampling to resolve all important frequency structure
in the radiation field at different depths.  This was achieved
through an adaptive frequency mesh of typically 10,000 grid
points and spectral resolution up to $10^6$, similar to the
resolution in the calculations of Abgrall et al.~(1992).
Another 3200 weaker lines were included using the equivalent
width approximation atop of the spectrum of the stronger lines.
Typically, these weak lines contribute less than 5\% to the
total pumping rate.  
The CPU-time-efficient transfer algorithm is
part of a fully time-dependent PDR code that will be described
in a forthcoming paper.

For comparison with the approximate treatment introduced
in \S\ref{sec:withoverlap}, the $\HH$ level populations were assumed to be
given by LTE at excitation temperature $T_{exc}$.  
Rather than examine the actual self-shielding function, we consider
the suppression of the dissociation rate, $\zeta_{diss}/\zeta_{diss}(0)$.
We compare the dissociation rate obtained from our
``exact'' radiative transfer equation with $\zeta_{diss}$
obtained from eq.\ (\ref{eq:withoverlap}), for $T_{exc}=100\K$ (Figure 1)
and $T_{exc}=300\K$ (Figure 2).  We see that the agreement is
excellent at all column densities $N_2$.
Agreement is of course expected for low column densities
$N_2\ltsim10^{18}\cm^{-2}$ where line overlap effects are
negligible (so that treatment of the pumping as due to isolated
lines is valid).  This agreement persists, however, at column
densities $N_2>10^{20}\cm^{-2}$ where line overlap effects cause
$f_{shield}$ to decline more rapidly than $N_2^{-0.5}$; the
agreement is excellent even out to the largest column densities
considered, $N_2=2\times10^{21}\cm^{-2}$, where line overlap
effects suppress the pumping rates by a factor $\sim10$.  We
have therefore validated the use of eq.\ (\ref{eq:withoverlap})
for subsequent calculations, in which we
solve for the non-LTE $\HH$ level populations.

\begin{figure}[t]
\epsscale{0.60}%size1
\plotone{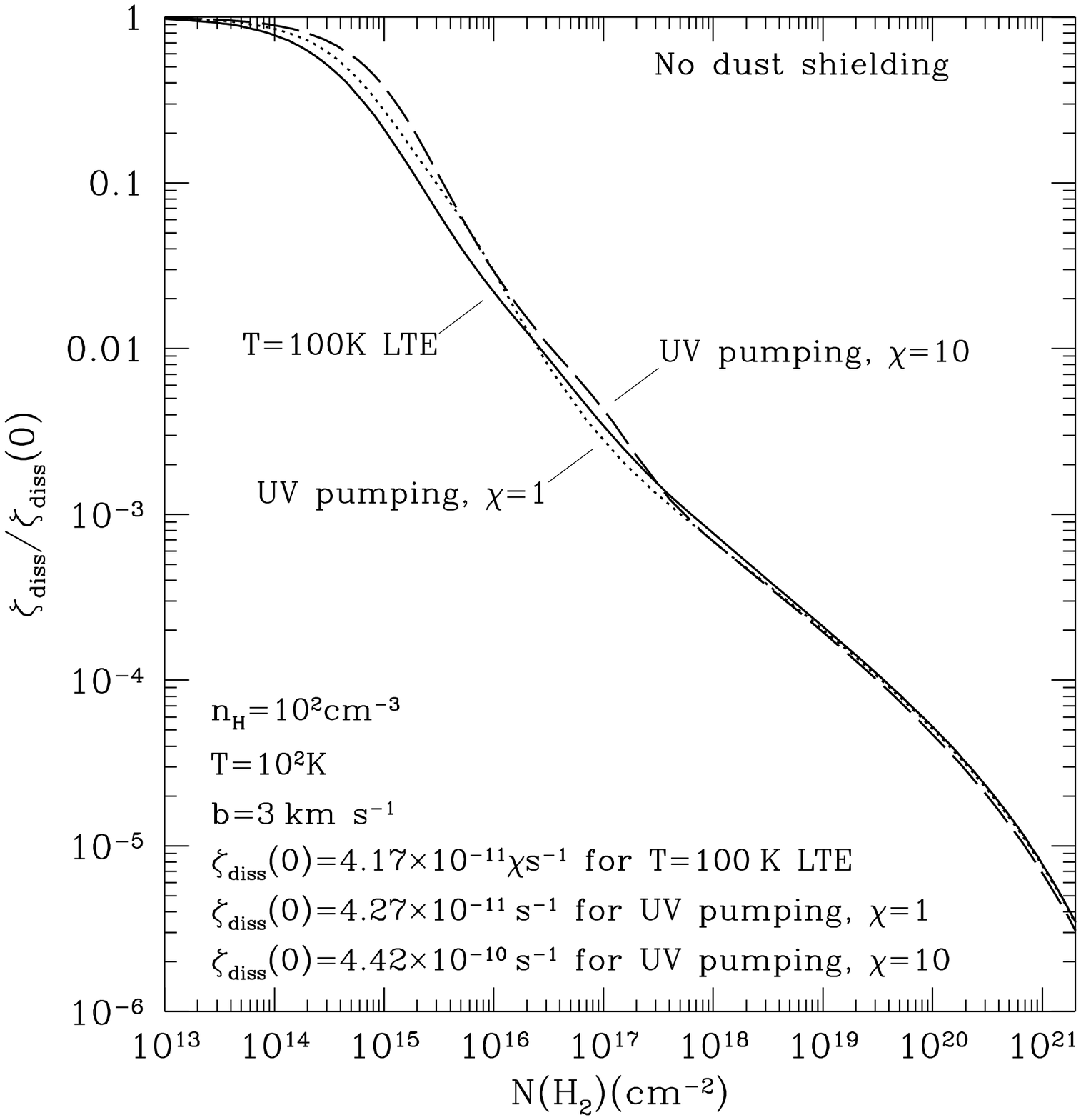}
\caption{
	Self-shielding for 3 photodissociation fronts: $T=100\K$
	LTE level populations (solid line),
	and UV pumping with $\chi=1$ (dotted) and $\chi=10$ (broken)
	(with $\nH=10^2\cm^{-3}$ and $T=10^2\K$).
	It is seen that the three self-shielding factors are essentially identical
	for $N(\HH)\gtsim 2\times10^{17}\cm^{-2}$.
	\label{fig:comparef}
	}
\end{figure}
\begin{figure}[t]
\epsscale{0.60}%size1
\plotone{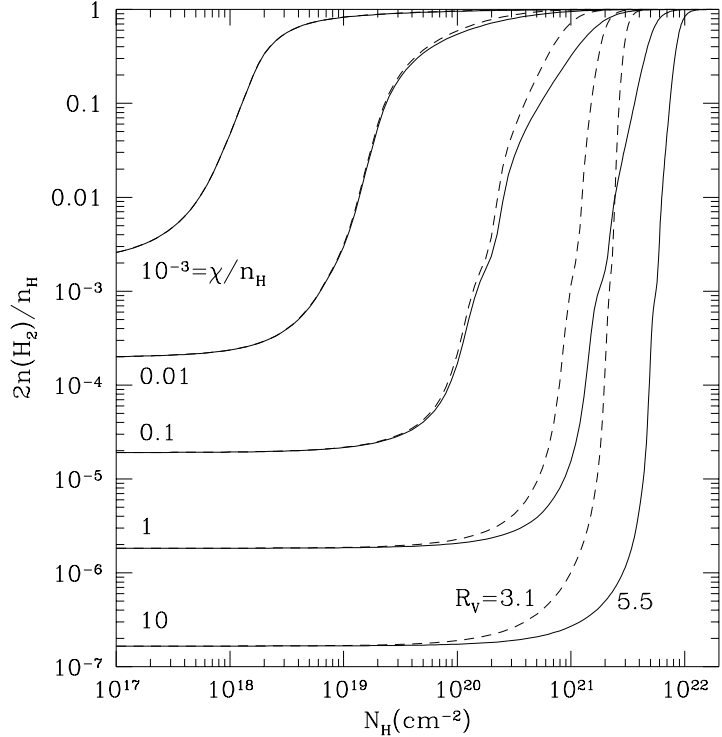}
\caption{
	$\HH$ fractions in stationary, plane-parallel
	photodissociation fronts for $\nH=10^2\cm^{-3}$ and
	$T_0=200\K$,
	for selected values of $\chi/\nH$ ($\cm^3$) and dust with
	$R_V=3.1$ ($\sigma_{d,1000}=2\times10^{-21}\cm^2$) and 
	$R_V=5.5$ ($\sigma_{d,1000}=6\times10^{-22}\cm^2$).
	$\lambda > 912\Angstrom$ radiation with $u_\nu\propto\nu^{-2}$
	is propagating in the $+x$ direction at $\NH=0$.
	$\NH$ is the total column density of H nucleons.
	Self-shielding of the $\HH$ is computed for 27983 lines
	using eq.\ (\protect\ref{eq:withoverlap})
	with $W_{max}=0.2$.
	\label{fig:fh2vsNH}
	}
\end{figure}
\begin{figure}[t]
\epsscale{0.60}%size1
\plotone{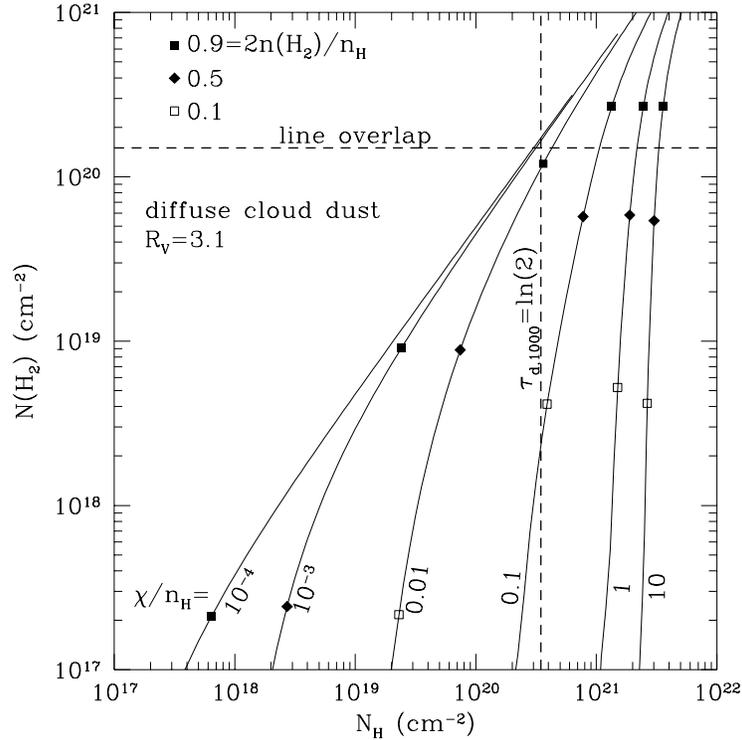}
\caption{
	$N(\HH)$ vs. $\NH$ for photodissociation fronts with
	diffuse cloud dust.
	Models are labelled by $\chi/\nH (\cm^3)$.
	Calculations were done for $\nH=10^2\cm^{-3}$, $T_0=200\K$.
	Line overlap contributes more than a factor of
	two to the self-shielding above the horizontal line.
	To the right of the vertical line dust shielding contributes more than a
	factor of 2.
	Points on each curve show locations where the $\HH$ fraction is 
	0.1, 0.5, and 0.9.
	It is seen that for diffuse cloud dust properties dust shielding is 
	important whenever line overlap occurs.
	\label{fig:NH2vsNH,diff.}
	}
\end{figure}
\begin{figure}[t]
\epsscale{0.60}%size1
\plotone{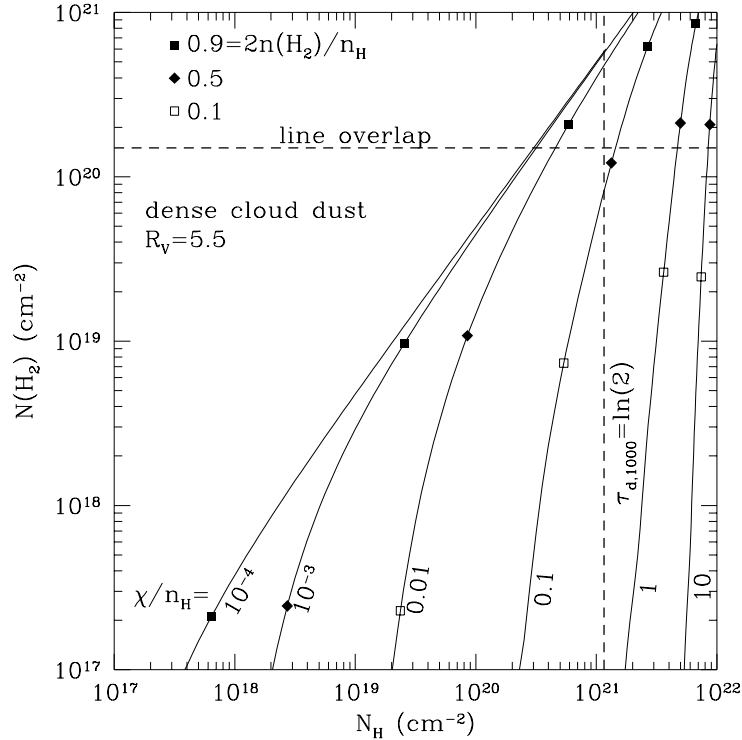}
\caption{
	Same as Fig.\ 
	\protect\ref{fig:NH2vsNH,diff.}, but for dense cloud dust.
	It is seen that line overlap can be important even when dust absorption
	plays a minor role (e.g., $\chi/\nH=.01\cm^3$ and 
	$N_\H=5\times10^{20}\cm^{-2}$).
	Calculations were done with $\nH=10^2\cm^{-3}$, $T_0=200\K$.
	\label{fig:NH2vsNH,dense}
	}
\end{figure}
\begin{figure}[t]
\epsscale{0.60}%size1
\plotone{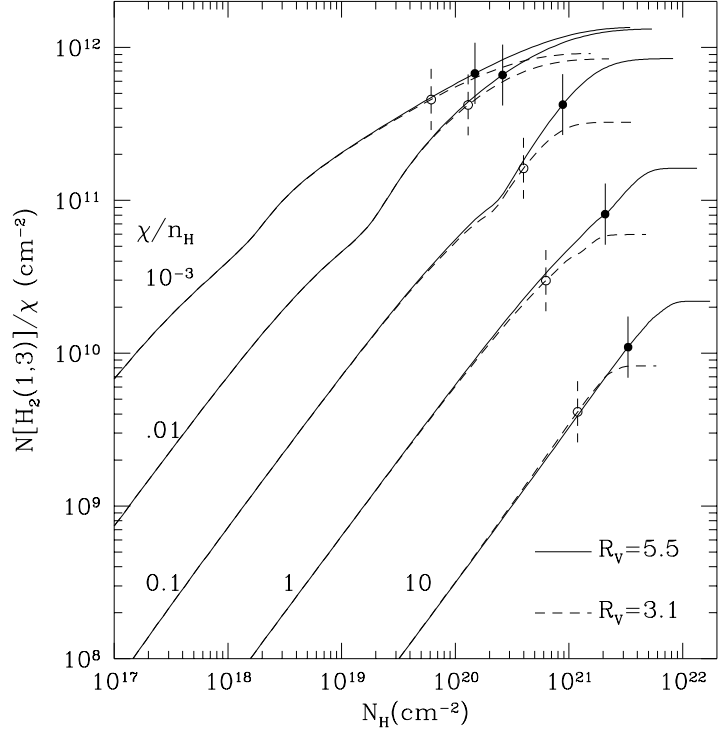}
\caption{
	Column density of $\HH(v=1,J=3)$ divided by $\chi$ 
	in stationary, plane-parallel photodissociation
	fronts, labelled by $\chi/\nH$ ($\cm^3$), for dust with $R_v=3.1$ 
	and 5.5 .
	$\NH$ is the total column density of H nucleons. 
	Self-shielding of the $\HH$ is computed using 
	eq.\ (\protect\ref{eq:withoverlap}) 
	with $W_{max}=0.2$.
	Calculations were done for $\nH=10^2\cm^{-3}$, $T_0=200\K$.
	Results are valid for $\chi \ltsim 2000$ 
	and $\nH \ltsim 10^5\cm^{-3}$ (so that UV pumping and collisional 
	deexcitation out of vibrationally-excited
	levels are slower than spontaneous decay).
	For each case a vertical line indicates the ``median''
	location for 1--0S(1) emission (see text).
	\label{fig:Nvj13}
	}
\end{figure}
\begin{figure}[t]
\epsscale{0.60}%size1
\plotone{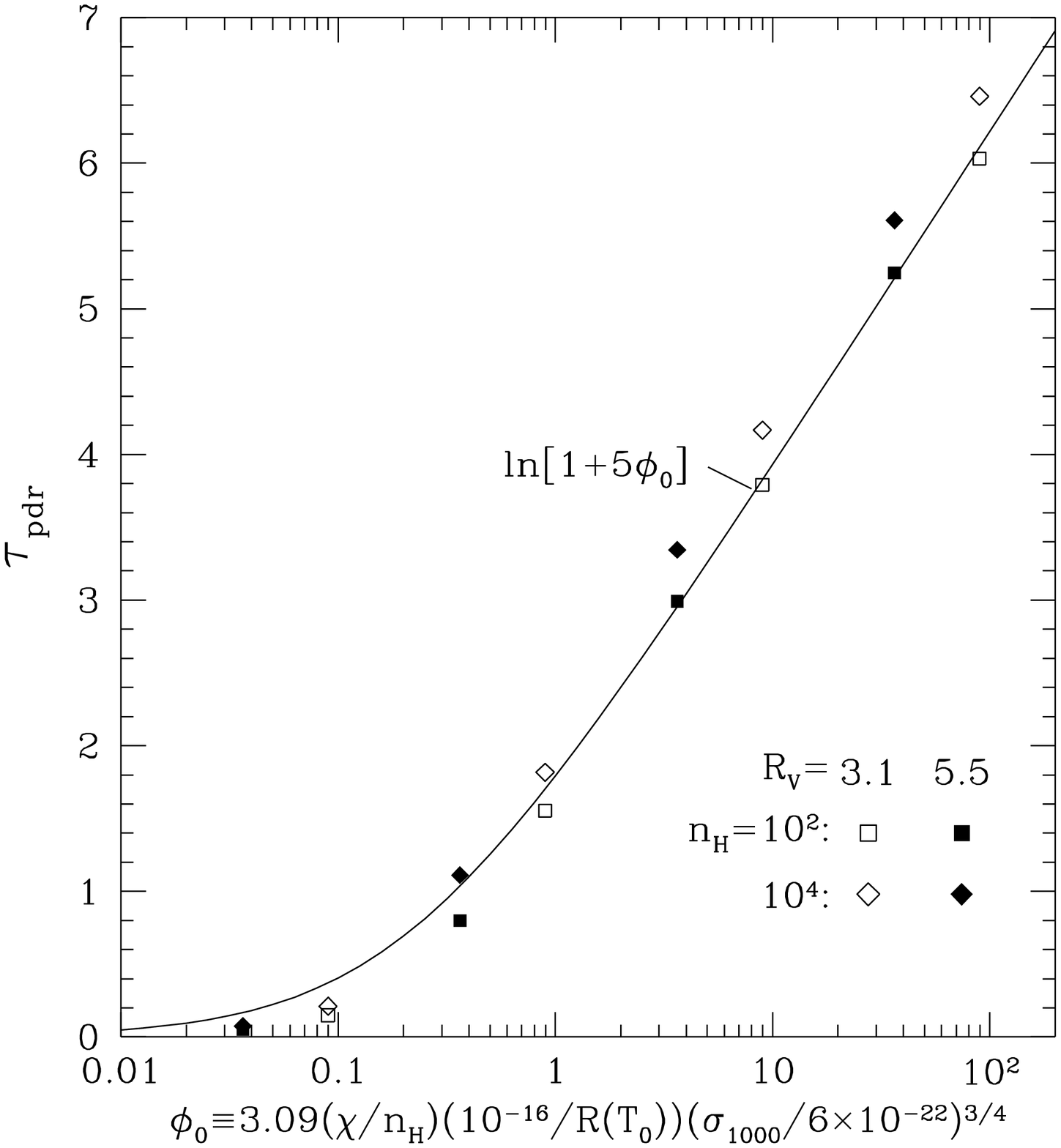}
\caption{
	The dust optical depth $\tau_{pdr}$ at $\lambda=1000\Angstrom$
	between the ionization front and the point where $2n(\HH)=n(\H)$,
	as a function of the dimensionless parameter $\phi_0$.
	Results are shown for two different extinction laws.
	It is seen that $\tau_{pdr}$ may be approximated by
	$\tau_{pdr}\approx\ln(1+
	5\phi_0)$.
	Numerical results are for $T_0=200\K$.
	\label{fig:taupdr}
	}
\end{figure}

For comparison with our multiline calculations,
in Figures \ref{fig:LTE100} and \ref{fig:LTE300} we have plotted the 
self-shielding
function for a single line in the approximation of Federman et al.
(1979), with the two parameters $r=1.3\times 10^{-3}$ and $N_2/\tau_D
= 10^{14.6}\cm^{-2}$ chosen to approximately match 
the total H$_2$ self-shielding curve for $T_{exc}=100$ K.
It is apparent that the self-shielding of a single line is
quite different from that of the total ensemble of overlapping H$_2$ lines.
The effects of line overlap were estimated by
de Jong, Dalgarno, \& Boland (1980); 
their result was used by Tielens \& Hollenbach (1985a) for
$N_2\gtsim2\times10^{15}\cm^{-2}$.
In Figures \ref{fig:LTE100} and \ref{fig:LTE300} we show the 
self-shielding function used by
Tielens \& Hollenbach.
The de Jong, Dalgarno, \& Boland formula evidently underestimates the
effects of line overlap, by approximately a factor of $\sim2.4$ at
$N(\HH)=10^{21}\cm^{-2}$.
Also shown in Figures \ref{fig:LTE100} and \ref{fig:LTE300} is the dissociation
probability $\langle p_{diss}\rangle$.
While $\langle p_{diss}\rangle$ varies from $0.10$ to $0.21$, it is generally
close to the average value of $\sim0.15$ (dotted line).

Figures \ref{fig:chi/nh=.01,T_0=200,Rv=3.1} and 
\ref{fig:chi/nh=.01,T_0=200,Rv=5.5} show the variation of the 
self-shielding factor
for dissociation, $\zeta_{diss}/\zeta_{diss}(0)$,
with depth in a cloud with $\chi/\nH=0.01\cm^3$,
for two extreme assumptions concerning the dust: diffuse cloud
dust with $R_V=3.1$ (Figure \ref{fig:chi/nh=.01,T_0=200,Rv=3.1}), 
and dense cloud dust with
$R_V=5.5$ (Figure \ref{fig:chi/nh=.01,T_0=200,Rv=5.5}). 
Figure \ref{fig:chi/nh=.1,T_0=200,Rv=5.5} shows $f_{shield}$ for dense
cloud dust, but with an enhanced radiation field
$\chi/\nH=0.1\cm^3$.  We see in Figures 
\ref{fig:chi/nh=.01,T_0=200,Rv=3.1}--\ref{fig:chi/nh=.1,T_0=200,Rv=5.5} 
that at $\HH$
column density $N_2=1\times10^{21}\cm^{-2}$, the $\HH$
self-shielding suppresses the photodissociation rate by a factor
$f_{shield}\approx6\times10^{-6}$.
It is also interesting to observe in 
Figs. \ref{fig:chi/nh=.01,T_0=200,Rv=3.1}--\ref{fig:chi/nh=.1,T_0=200,Rv=5.5}
the decline in $\langle p_{diss}\rangle$ at large values of $N_2$.
This decrease (not seen in the dustless models of Figs. 
\ref{fig:LTE100} and \ref{fig:LTE300})
is due to attenuation of the FUV by the dust, resulting in
``reddening'' of the radiation to which the $\HH$ is exposed,
decreasing the relative importance of the photoexcitations to higher levels
$u$, with larger values of $p_{diss,u}$.

Also plotted on Figures 
\ref{fig:chi/nh=.01,T_0=200,Rv=3.1}--\ref{fig:chi/nh=.1,T_0=200,Rv=5.5}
is the ``square-root approximation''
$f_{shield}$ $\approx$ $7\times10^{-5}(10^{20}\cm^{-2}/N_2)^{0.5}$;
it is evident that except over the limited range
$10^{17}\ltsim N_2\ltsim 10^{20}\cm^{-2}$
the $\HH$ self-shielding function is quite different
from the asymptotic $N_2^{-0.5}$ behavior which is often assumed
(e.g., Hill \& Hollenbach 1978; 
Sternberg 1988; Goldschmidt \& Sternberg 1995; Hollenbach \& Natta 1995).
The differences arise from two effects.  For $10^{15}\ltsim
N_2\ltsim 10^{17}\cm^{-2}$ the actual self-shielding function
declines with increasing $N_2$ much more rapidly than
$N_2^{-0.5}$ while many of the strongest
absorbing transitions become optically thick and enter
the ``flat'' portion of the curve-of-growth.  The
column density range of this steep drop is wider than that of a
single line because initially less important transitions (i.e.
those with smaller oscillator strength and/or smaller
populations) are still optically thin and 
increase in relative importance when the stronger lines become saturated.
Only at $N_2> 10^{17}\cm^{-2}$ have
all important pumping lines reached the square-root part of the
curve-of-growth.
For $N_2\gtsim10^{20}\cm^{-2}$
$f_{shield}$ falls off more rapidly than $N_2^{-0.5}$ as the
result of line overlap.
The function $7\times10^{-5}(10^{20}\cm^{-2}/N_2)^{0.5}$ underestimates
$f_{shield}$ by a factor $\sim20$ at $N_2=10^{15}\cm^{-2}$, and overestimates
$f_{shield}$ by a factor $\sim 4$ at $N_2=10^{21}\cm^{-2}$.
 
The importance of line overlap is shown by Figure \ref{fig:eqwidths}, 
where we plot
the total equivalent width $W$ summed over Lyman and Werner series for
$\HH$, plus the H Lyman series from Lyman$\beta$ through H$1-15$.
Treating the frequencies of the lines in the $1110-912\Angstrom$ band
as independent random variables, our statistical treatment of line overlap,
with $f_{shield}\propto\exp(-W/0.2)$, gives a factor of 2 suppression
of $f_{shield}$ 
when the total equivalent
width $W=0.2\ln(2)=0.14$, which occurs for
$N_2\approx3\times10^{20}\cm^{-2}$ (see Fig. \ref{fig:eqwidths})
 
If a power-law fit to the self-shielding factor $f_{shield}$ is desired,
the $N_2^{-3/4}$ power law of equation (\ref{eq:powfit}), shown in Figures 
\ref{fig:chi/nh=.01,T_0=200,Rv=3.1}--\ref{fig:chi/nh=.1,T_0=200,Rv=5.5},
seems most suitable; it succeeds in reproducing
$f_{shield}$ to 
within a factor of 2 for $10^{14}< N_2 < 5\times10^{20}\cm^{-2}$.
 
If a more accurate approximate representation of $f_{shield}$ is desired,
we favor equation (\ref{eq:goodfit}), also shown in Figures 
\ref{fig:chi/nh=.01,T_0=200,Rv=3.1}--\ref{fig:chi/nh=.1,T_0=200,Rv=5.5}.
This functional form does an excellent job in reproducing the initial rapid
decline in $f_{shield}$ for $10^{14}<N_2<10^{17}\cm^{-2}$, the ``square-root''
behavior for $10^{17}<N_2<10^{20}\cm^{-2}$, and the rapid fall-off due to
line overlap for $N_2 > 10^{20}\cm^{-2}$.
 
As already seen from comparison of Figures \ref{fig:LTE100} and 
\ref{fig:LTE300},
the detailed dependence of $f_{shield}$ on $N_2$ depends upon the 
rovibrational distribution of
the absorbing $\HH$.
In Figure \ref{fig:comparef} 
we compare $f_{shield}$ computed for: 
(1) an LTE ($T_{exc}=100\K$) $\HH$ level distribution
(see Figure \ref{fig:LTE100});
(2) a non-LTE rovibrational distribution due to UV pumping with $\chi=1$
and collisional excitation/deexcitation with $\nH=10^2\cm^{-3}$ 
and $T_0=200\K$ (see Figure \ref{fig:chi/nh=.01,T_0=200,Rv=5.5});
and (3) same as (2), but with $\chi=10$
(see Figure \ref{fig:chi/nh=.1,T_0=200,Rv=5.5}).
In order to bypass any questions related to dust opacities, the
plots in Figure \ref{fig:comparef} are for cases with zero dust extinction.
We see that for $10^{14}\ltsim N_2\ltsim 10^{18}\cm^{-2}$ $f_{shield}$
does show a dependence on the details of the $\HH$ rovibrational distribution.
When there is higher excitation,
the pumping occurs via a larger number of lines, so that
the $\HH$ remains optically-thin a bit longer,
and self-shielding is less effective until one enters the heavily damped
regime at $N_2\gtsim 10^{18}\cm^{-2}$.
Overall, however, the self-shielding function $f_{shield}(N_2)$ is
relatively insensitive to the effects of UV pumping with 
$\chi\ltsim10^3$.
 
Figure \ref{fig:fh2vsNH} shows profiles of the fractional 
abundance of $\HH$ as a function of
total column density $\NH$ for five different values of $\chi/\nH$ 
($=10^{-3}$, 0.01, 0.1, 1, and 10 $\cm^3$) and
two different dust opacity laws ($R_V=3.1$ and 5.5).
As the radiation intensity
to density ratio
$\chi/\nH$ is raised, the column
density of atomic, dusty gas increases.
 
In Figures \ref{fig:NH2vsNH,diff.} and \ref{fig:NH2vsNH,dense} 
we show the run of
$N(\HH)$ vs.\ $\NH$ for models parameterized by
$\chi/\nH$.
Models assuming diffuse cloud dust properties are shown in 
Fig.\ \ref{fig:NH2vsNH,diff.};
dense cloud dust is considered in Fig.\ \ref{fig:NH2vsNH,dense}.
The region in which line overlap is important is shown in each figure,
as is the region where attenuation of the UV by dust is significant.
It is seen from Fig.\ \ref{fig:NH2vsNH,dense} 
that line overlap plays an important role in
self-shielding.
For example, a dense cloud dust model with $\chi/\nH=0.01\cm^3$
has line overlap suppressing the dissociation rate by factors of 2-4
at column densities $5\ltsim \NH \ltsim 12\times10^{20}\cm^{-2}$
where dust extinction attenuates the radiation field by less than a
factor of two.

The 1--0S(1) line originates from $\HH(v=1,J=3)$.
In Figure \ref{fig:Nvj13} we show the column density of this excited state,
divided by $\chi$, as a function of the total column density $\NH$,
so that one can see how deep into the cloud one must go to account for
most of the
1--0S(1) line emission.
For each case we indicate the ``median'' for 1--0S(1) emission: the
location where 50\% of the 1--0S(1) emission occurs on either side.
For $\chi/\nH=10^{-3}\cm^3$, 
the 1--0S(1) median is at 
$\NH=1.5\times10^{20}\cm^{-2}$, but for
$\chi/\nH=0.1\cm^3$, for example, the 1--0S(1) median is at
$\NH=8.5\times10^{20}\cm^{-2}$ (for $R_V=5.5$ dust).

\subsection{Width of the PDR}

To assess the importance of shielding by dust, it is useful to
define the dust optical depth of the PDR, $\tau_{pdr}$, which we take
to be the dust optical depth at $\lambda=1000\Angstrom$ measured to the
point where 50\% of the local hydrogen is molecular: $2n(\HH)/\nH=0.5$.
In Figure \ref{fig:taupdr}
we plot $\tau_{pdr}$ as a function of the
parameter $\phi_0$ defined in Paper I:
\beq
\phi_0 \equiv 25.4 \left({\chi\, \cm^{-3} \over \nH}\right)
\left({3\times10^{-17}\cm^3\s^{-1} \over R(T_0)}\right)
\left({\sigma_{d,1000}\over 2\times10^{-21}\cm^2}\right)^{3/4}~,
\label{eq:phi_0}
\eeq
where the rate coefficient $R$ for $\HH$ formation on grains is
evaluated at temperature $T_0$.
For $R$ given by eq.\ (\ref{eq:h2form}) this becomes
\beq
\phi_0 = 5.15 \left({\chi\cm^{-3}\over\nH}\right)
\left({100\K\over T_0}\right)^{1/2}
\left({\sigma_{d,1000}\over 6\times10^{-22}\cm^2}\right)^{3/4}~.
\eeq
We see that shielding by dust is important ($\tau_{pdr}>1$) for 
$\phi_0>0.5$,
but relatively unimportant for $\phi_0<0.1$.
A good fit to the numerical results is provided by the simple
fitting function
(cf. Fig. \ref{fig:taupdr})
\beq
\tau_{pdr}\approx \ln(1+5\phi_0)~~~.
\label{eq:taupdr}
\eeq
\begin{figure}[t]
\epsscale{0.60}%size1
\plotone{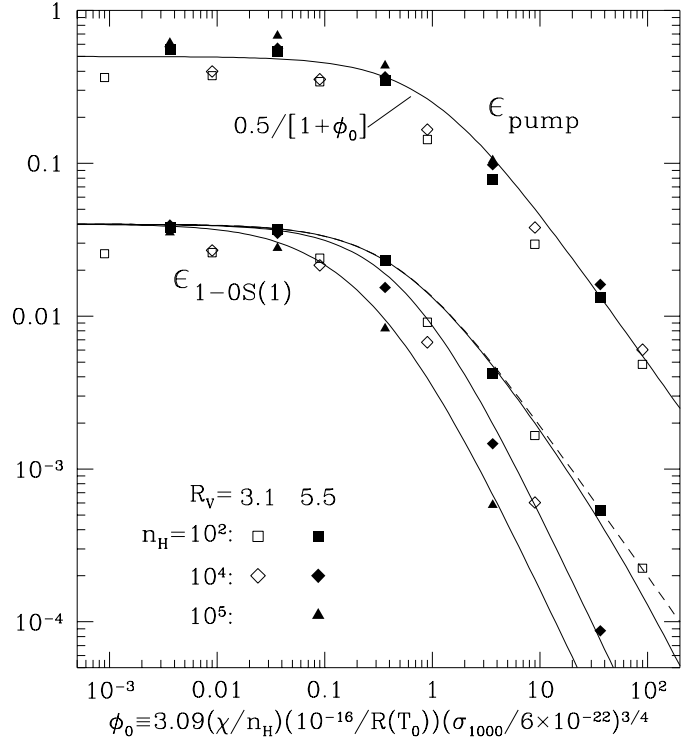}
\caption{
	The UV pumping efficiency $\epsilon_{pump}$, the fraction of incident
	$1110-912\Angstrom$ photons which are absorbed by $\HH$ rather than
	dust,
	and $\epsilon_{1-0S(1)}$, the number of 1--0S(1) photons emitted per
	incident $1100-912\Angstrom$ photon.
	Numerical results are for $T_0=200\K$.
	The estimate (\ref{eq:eps_pump_approx})
	for $\epsilon_{pump}$ is shown as a solid curve.
	The estimate (\ref{eq:epsilon10s1fit}) for $\epsilon_{1-0S(1)}$
	is shown for dense cloud dust with 
	$R(T_0)=8.46\times10^{-17}\cm^3\s^{-1}$
	for $\nH\rightarrow0$ (broken curve)
	and for $\nH=10^2$, $10^4$, and $10^5\cm^{-3}$ (solid curves).
	\label{fig:epsilons}
	}
\end{figure}

\subsection{Pumping efficiency\label{pump_eff}}

Because of the increased importance of dust for large $\phi_0$, the
efficiency of conversion of incident UV into $\HH$ fluorescence is
reduced.
One measure of this is the ``pumping efficiency'', the ratio
\beq
\epsilon_{pump} = {\int \zeta_{pump} dN_2 \over F}
\label{eq:epspumpdef}~~~,
\eeq
where $F$ is the incident flux of 1110--912\AA\ photons 
(cf. eq.\ \ref{eq:fdef}).
Since there is only a minor contribution to pumping by $\lambda>1110\Angstrom$
photons, 
$\epsilon_{pump}$ is essentially the fraction of the 
incident $1110-912\Angstrom$
photons which are absorbed by $\HH$, rather than dust, so that the
$\HH$ dissociation rate/area and (for $\nH\ltsim10^4\cm^{-3}$)
the fluorescent surface brightness of the cloud are proportional to
$\epsilon_{pump} F\propto\epsilon_{pump}\chi$.
 
In Figure 
\ref{fig:epsilons} we plot $\epsilon_{pump}$ versus the parameter $\phi_0$,
for models with $T_0=200\K$ but a variety of densities and both diffuse cloud
dust and dense cloud dust.
We see that $\epsilon_{pump}\approx0.5$ for $\phi_0\ltsim0.1$.
Strictly speaking, $\epsilon_{pump}$ is a function of both $\chi/\nH R$ and
$\sigma_d$,\footnote{
	It is clear, for example, that for $\sigma_d\rightarrow0$
	(in which case $\phi_0\rightarrow0$) we should 
	have $\epsilon_{pump}\approx 1$
	since all 1110--912\AA\ photons must eventually be absorbed by
	$\HH$ (neglecting the H Lyman lines and CI absorption).
	However, for values of $\sigma_d$ and $\phi_0$ of practical
	interest, we find that $\epsilon_{pump}\approx0.5$ for $\phi_0\ll1$.
	}
but we find that the pumping efficiency $\epsilon_{pump}$ can in practice
be approximated as depending on the single parameter $\phi_0$,
with the fitting function
\beq
\epsilon_{pump} \approx {0.5\over 1 + \phi_0}
\label{eq:eps_pump_approx}
\eeq
providing a reasonable fit to our results (cf. Fig.\ \ref{fig:epsilons}).
Also shown in Fig.\ \ref{fig:epsilons} 
is the ``efficiency'' $\epsilon_{1-0S(1)}$ for emission of the
strong 1--0S(1) line.  We define this in terms of the surface brightness
$I(1\rightarrow0{\rm S}(1),\theta=0)$ measured normal to the PDR
(and including the effects of internal dust; see eq.\ [\ref{eq:S_lu}]):
\beq
\epsilon_{1-0S(1)}\equiv {4\pi I(1\!\rightarrow\!0{\rm S}(1),\theta=0) \over hc/\lambda_{1-0S(1)}} 
{1\over F} ~~~.
\label{eq:epsilon10s1}
\eeq
Provided that $\nH\ltsim10^5\cm^{-3}$ (so that collisional deexcitation
does not compete with spontaneous decay out of (1,3))
we find that $\epsilon_{1-0S(1)}$ may be approximated by
\beq
\epsilon_{1-0S(1)}\approx {0.04\over 1+2\phi_0}~
{1\over [1+(\chi/2000)^{0.5}]} ~~~,
\label{eq:epsilon10s1fit}
\eeq
showing that for $\phi_0\ltsim0.1$ and $\chi\ltsim2000$,
$\epsilon_{1-0S(1)}\approx0.08\epsilon_{pump}$:
$\sim$8\% of the UV pumping events 
result in fluorescent
emission of a 1--0S(1) photon.
Thus, for $\nH\ltsim10^5\cm^{-3}$,
\beq
I(1\!\rightarrow\!0{\rm S}(1),\theta=0)\approx {3.6\times10^{-8}\chi\over
(1+2\phi_0)
[1+(\chi/2000)^{0.5}]}
\erg\cm^{-2}\s^{-1}\sr^{-1} ~~~.
\eeq

Unlike $\epsilon_{pump}$, we see from Fig.\ \ref{fig:epsilons}
that $\epsilon_{1-0S(1)}$ is not determined solely by $\phi_0$, but also
shows sensitivity to the actual values of $\nH$ and $\chi$ when
$\nH\gtsim10^5\cm^{-3}$ or $\chi\gtsim10^3$.
This is because at high densities or high intensities, (1) the rotational
population is changed, affecting the probability of populating the
$v=1,J=3$ level, and (2) collisional deexcitation (at $\nH\gtsim10^5\cm^{-3}$)
or radiative pumping (at $\chi\gtsim10^3$) may act to depopulate the
$v=1,J=3$ level.
For $\nH\ltsim10^5\cm^{-3}$ we find that eq.\ (\ref{eq:epsilon10s1fit})
provides a good estimate for the intensity of the 1--0S(1) line.

\section{Fluorescent Emission Spectra}
\subsection{Effects of Internal Extinction}

The true column density in level $(v,J)$
along a line making an angle $\theta$ with respect
to the normal is simply
\beq
N(v,J) = {1\over\cos(\theta)}\int_0^\infty dx \, n(v,J)~~~.
\eeq
The emission in vibration-rotation lines of $\HH$ is
\beq
I(u\!\rightarrow\!l,\theta)
= {1\over\cos(\theta)}
\int_0^\infty dx \, 
{A_{lu}\over4\pi}n_u{hc\over \lambda_{ul}}
\exp[-\NH(x) \sigma(\lambda_{ul})/\cos(\theta)]~,
\label{eq:S_lu}
\eeq
where $\theta$ is the angle of inclination of the front as
seen by the observer, and
we have allowed for absorption of the fluorescent emission
by dust within the photodissociation front 
($\NH(x)=\int_0^x dx^\prime
\nH(x^\prime)$).

When a surface brightness $I_{obs}(v,J\rightarrow v^\prime,J^\prime)$
is observed in an $\HH$ transition $v,J\rightarrow v^\prime,J^\prime$,
one may directly
compute the ``apparent'' column density $N_{app}(v,J)$ in the
upper level from
\beq
N_{app}(v,J)={4\pi\lambda_{ul} I_{obs}(v,J\rightarrow v^\prime,J^\prime)\over
hc A(v,J\rightarrow v^\prime,J^\prime)}~.
\label{eq:napp}
\eeq
Because of extinction, however, the true column density is larger.
It is usual to infer an ``observed'' or ``dereddened'' column density
\beq
N_{obs}(v,J)=N_{app}(v,J)~10^{0.4A_\lambda}~,
\label{eq:nobs}
\eeq
where $A_\lambda$ is the estimated extinction, in magnitudes, 
at the wavelength $\lambda$
of the observed transition $v,J\rightarrow v^\prime,J^\prime$.
\begin{figure}[t]
\epsscale{0.56}%size2
\plotone{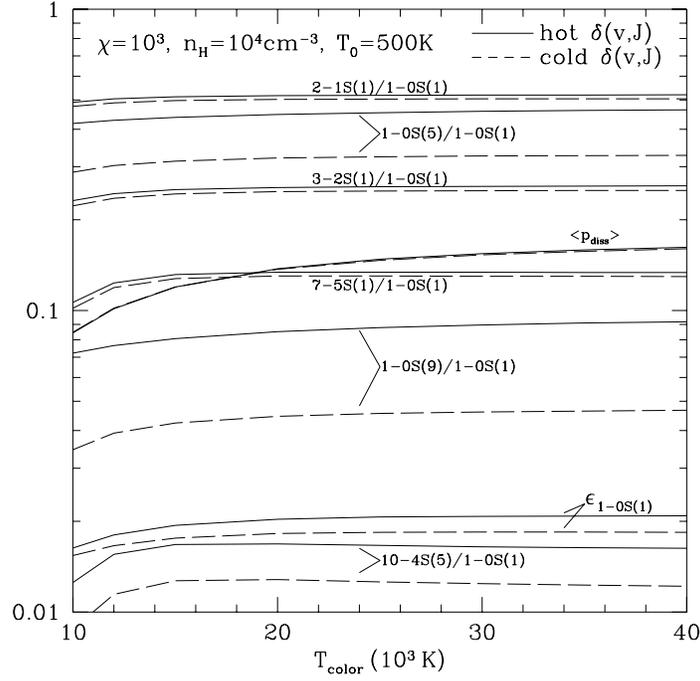}
\caption{
	Selected line intensity ratios for 
	$\chi\!=\!10^3$, $\nH\!=\!10^4\cm^{-3}$,
	versus the color temperature $T_{color}$ of the incident
	radiation field.
	Curve labelled $\epsilon_{1-0S(1)}$ is the
	efficiency for 1--0S(1) line emission
	[eq.(\protect{\ref{eq:epsilon10s1}})].
	Solid and broken curves are, respectively, 
	for newly-formed $\HH$ which is 
	rovibrationally-``hot''
	($T_{\rm f}\!=\!5\times10^4\K$) and
	``cold''($T_{\rm f}\!=\!2000\K$).
	The PDR is assumed to have
	$T_0=500\K$, dust with $R_V=5.5$, 
	$\sigma_{d,1000}=6\times10^{-22}\cm^2$,
	and $\HH$ formation rate given by eq.\ (\protect{\ref{eq:h2form}}).
	Line ratios are seen to be essentially independent of $T_{color}$ for
	$T_{color}\gtsim12000\K$.
	Intensities of lines out of levels ($v$,$J$) with 
	$J=3$ [S(1) lines] are essentially independent of the
	value of $T_{\rm f}$.
	Intensities of lines out of levels with $J=7$ [S(5) lines] are
	increased by $\sim50\%$ when $T_{\rm f}$ is varied from
	$2000\K$ to $5\times10^4\K$.
	Intensities of lines out of levels with $J=11$ [S(9) lines] are
	a factor of $\sim2$ stronger when the newly-formed $\HH$ is
	``hot'' rather than ``cold''.
	\label{fig:tcolor_and_tform}
	}
\end{figure}

\subsection{Sensitivity of the PDR to Incident Spectrum 
	\label{sec:incident_spectrum}}

PDRs may be produced by radiation from stars ranging from late B-type to
early O-type, and the properties of the PDR will in principle depend
upon the color temperature of the illuminating radiation as well as
on $\chi$, measuring the intensity at $1000\Angstrom$.
Fig.\ref{fig:tcolor_and_tform} shows $\epsilon_{1-0S(1)}$ and
selected line ratios for a series of PDR models where $\chi$ and $\nH$
are held constant but the illuminating
radiation is taken to be a dilute blackbody with color temperature
$T_{color}=10000\K$ (A0 spectrum) to
$T_{color}=40000\K$ (O4 spectrum).
The models have $\chi=10^3$ and $\nH=10^4\cm^3$ -- for this density
collisional deexcitation is relatively unimportant for
vibrationally-excited levels, so that the emission from these
models should show maximum sensitivity to the spectrum of the
illuminating radiation.
We see, however, that the emission properties of the PDR are essentially
independent of $T_{color}$ for $T_{color}>12000\K$.
This insensitivity allows us to limit our study to models computed for
power-law spectra with $\alpha=-2$ (corresponding to $T_{color}=29000\K$)
and apply them to PDRs produced by a broad range of star types.

\subsection{Sensitivity to $\hhini(v,J)$ for Newly-Formed $\HH$
	\label{sec:phi_0}}
In Fig.~\ref{fig:tcolor_and_tform} we also investigate the sensitivity
of the emission spectrum to the assumed distribution function
$\hhini(v,J)$ of newly-formed $\HH$.
The solid curves are for models with ``hot'' $\HH$, with 
$T_{\rm f}=5\times10^4\K$, and the broken curves are for models with
``cold'' $\HH$, with $T_{\rm f}=2000\K$.
We see that the intensities of lines out of levels with low $J$
[e.g., S(1) lines out of $J=3$] are essentially unaffected by the
assumed properties of $\hhini$, even for high vibrational levels
[e.g., 7--5S(1)].
Lines out of levels with higher $J$ values show increasing sensitivity
to the assumed $\hhini$:
S(5) lines out of $J=7$ levels vary by about 40\% between models with
hot or cold $\HH$,
and S(9) lines out of $J=11$ levels are about a factor of $\sim2$ stronger
in models with hot $\HH$ versus cold $\HH$.

It is interesting to note that the sensitivity to the assumed $\hhini(v,J)$
is primarily determined by the $J$ value of the upper state, and
seems not to depend particularly on the value of $v$: the
10--4S(5) line and the 1--0S(5) line each increase by about 50\% as
the newly-formed $\HH$ is changed from ``cold'' to ``hot''.
Unfortunately, as we show below, the rotational distribution of the 
$\HH$ depends on $\chi$, $\nH$, and the gas temperature;
unless these other properties of the PDR are known quite accurately,
it does not seem likely that we will be able to use observations
of PDR emission spectra to determine the distribution function
$\hhini(v,J)$ of newly-formed $\HH$.

\subsection{1--0S(1)/2--1S(1) line ratio\label{sec:lineratio}}

\begin{figure}[t]
\epsscale{0.60}%size1
\plotone{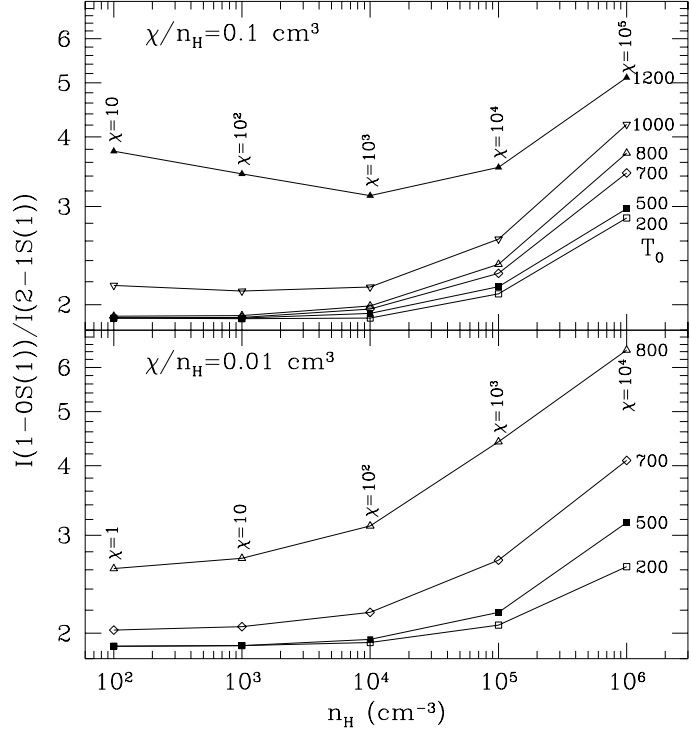}
\caption{
	1--0S(1)/2--1S(1) intensity ratio vs.\ density $\nH$.
	The dust is assumed to have $R_V=5.5$, 
	$\sigma_{d,1000}=6\times10^{-22}\cm^2$,
	and $R(T)$ from eq.\ (\protect{\ref{eq:h2form}}).
	Curves are labelled by $T_0$ (see eq.\ \protect{\ref{eq:tprof}}).
	For pure fluorescence ($\nH\ll10^5\cm^{-3}$ and $T_0\ll500\K$)
	we have 1--0S(1)/2--1S(1)=1.9;
	collisional effects increase the ratio by collisional
	excitation of $v=1$, and preferential deexcitation of
	$v=2$.
	\label{fig:10S1/21S1}
	}
\end{figure}

The 1--0S(1)/2--1S(1) line ratio is frequently used to characterize $\HH$
line-emitting regions.  In Fig.\ \ref{fig:10S1/21S1}
we show this intensity ratio (for PDRs viewed face-on) as a function
of gas density $\nH$, and a number of values of the temperature $T_0$
[see eq.\ (\ref{eq:tprof})].
Results are shown for two different values of the ratio $\chi/\nH$.
At low densities and temperatures the line ratio 
(for $u_\nu\propto\nu^{-2}$, or $T_{color}\approx3\times10^4\K$) 
is $I$(1--0S(1))/$I$(2--1S(1))=1.9, reflecting
excitation by radiative pumping, with negligible collisional
deexcitation (or radiative excitation) out of vibrationally-excited levels.
As seen from Fig.\ref{fig:tcolor_and_tform}, the line ratio is insensitive
to the value of $T_{color}$.
\begin{figure}[t]
\epsscale{0.60}%size1
\plotone{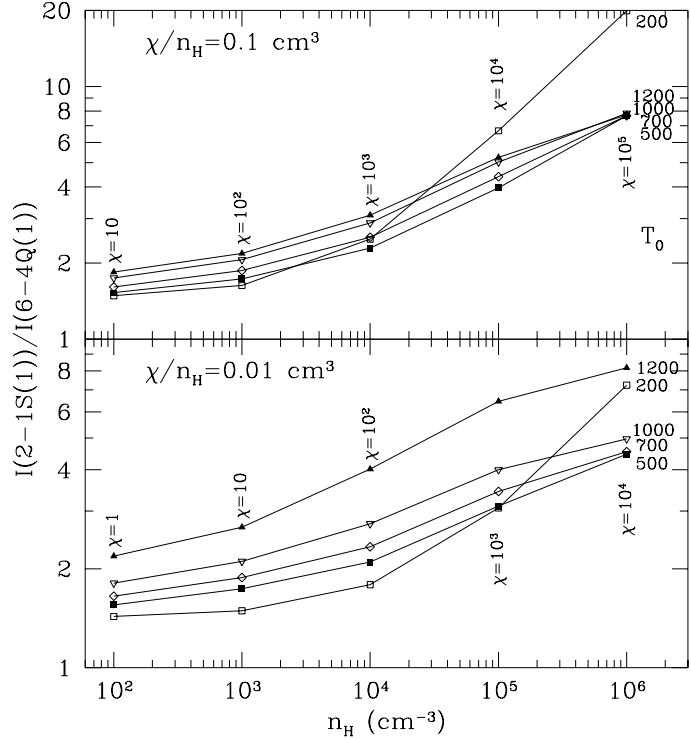}
\caption{
	As in Fig.\ \protect{\ref{fig:10S1/21S1}}, but for
	the 2--1S(1)/6--4Q(1) intensity ratio.
	The line ratio is affected by the rotational distribution
	of the $\HH$ (which explains why the line ratio depends
	upon $T_0$ even for $\nH=10^2\cm^{-3}$, and upon the
	effects of collisional deexcitation of $(v,J)=(2,3)$ and
	$(6,1)$.
	\label{fig:21S1/64Q1}
	}
\end{figure}

As the temperature and density are increased, two effects occur:
collisional deexcitation and collisional excitation.
Collisional deexcitation out of the $v=2$ (and higher) levels 
acts to reduce the intensity of 2--1S(1) emission; while collisional
deexcitation also depopulates the $v=1$ levels, the rate coefficients for
deexcitation out of $v=2$ levels (by collisions with 
H; Martin \& Mandy 1995)
are significantly larger than those
for deexcitation out of $v=1$, hence collisional deexcitation tends to
raise the 1--0S(1)/2--1S(1) line ratio.

If the temperature is high enough, collisional excitation of the $v=1,J=3$
level can also become important.
For $\chi/\nH=0.01\cm^3$ and $\nH \ltsim 10^4\cm^{-3}$, collisional
excitation contributes $\sim5\%$ of the 1--0S(1) emission for
$T_0=700\K$, $\sim30\%$ of the 1--0S(1) emission for
$T_0=800\K$, and dominates the 1--0S(1) emission for $T_0\gtsim850\K$.

As $\chi/\nH$ increases, collisional excitation becomes
relatively less important (at fixed $T_0$) for two reasons:
First of all, our adopted temperature profile (eq.\ [\ref{eq:tprof}]) has
the highest temperature in the region where the $\HH$ abundance is low.
In the optically-thin region, the $\HH$ fraction is proportional to
$\nH/\chi$, so that increasing $\chi/\nH$ decreases the amount of $\HH$
present in the hottest part of the PDR (e.g., where $0.8T_0<T<T_0$).
Secondly, when $\chi/\nH$ is increased, $\tau_{pdr}$ 
increases (see Fig.\ \ref{fig:taupdr}).
For our assumed temperature profile (\ref{eq:tprof}), this implies a
reduction in $T$ at the location where $2n(\HH)=n(\H)$, thereby decreasing
the rate of collisional deexcitation.
Therefore, when $\chi/\nH$ is increased, collisional excitation of 1--0S(1)
does not become important until higher temperatures: for $\chi/\nH=0.1\cm^3$
we see that for $T_0=1000\K$ the collisional excitation contributes only
$\sim10\%$ of the 1--0S(1) emission, and only begins to dominate for
$T_0\gtsim1200\K$.

\subsection{2--1S(1)/6--4Q(1) line ratio\label{sec:21S1/64Q1}}

The 6--4Q(1) line ($\lambda=1.6015\micron$) 
is sufficiently strong for spectroscopic
measurement (see, e.g., Luhman \& Jaffe 1996),
and the 2--1S(1)/6--4Q(1) line ratio is
a useful indicator of conditions in the region where UV pumping
is taking place.
Whereas the 2--1S(1)/1--0S(1) line ratio can be affected by collisional
excitation of (1,3) ($E/k=6951\K$), the 2--1S(1)/6--4Q(1) line ratio is much less
sensitive to collisional excitation because the (2,3) level has
$E/k=12550\K$).
The 2--1S(1)/6--4Q(1) ratio is therefore a relatively direct probe of
collisional deexcitation of (6,1) vs.\ (2,3).

The dependence of the 2--1S(1)/6--4Q(1) line ratio on the density
$\nH$ and temperature profile parameter $T_0$ is shown in 
Fig.\ \ref{fig:21S1/64Q1}.
Because the
2--1S(1) line originates from $J=3$ while the
6--4Q(1) line originates from $J=1$, the line ratio depends on
the gas temperature even at low densities where collisional
deexcitation of the levels is negligible:
the line ratio increases with increasing $T_0$ at low densities.
When the gas density is high enough that collisional deexcitation
begins to compete with radiative decay of the $(v,J)=(6,1)$ level,
the line ratio increases, since the collisional rate coefficients
(Martin \& Mandy 1995) are such that the $(6,1)$ level is more
strongly affected than the $(2,3)$ level.
At the highest density $\nH=10^6\cm^{-3}$,
we obtain very high values of 2--1S(1)/6--4Q(1) when
$T_0=200\K$ (see Fig.\ \ref{fig:21S1/64Q1}).
At these high densities the relative strengths of 2--1S(1) and 6--4Q(1)
are primarily determined by the relative rates of collisional deexcitation
of $(2,3)$ and $(6,4)$ by atomic H.
We remind the reader that we use an uncertain extrapolation of the results 
of Martin \& Mandy (1995) to low temperatures (see eq.\ (\ref{eq:mmextrap}).
For the temperatures $T < 200\K$ in our $T_0=200\K$ models, these
extrapolated rates are {\it very} uncertain; 
we have included these models mainly
to show that extreme values of the 2--1S(1)/6--4Q(1) line ratio
are possible at high densities.
Reliable calculations of line ratios for high density, low temperature models 
will not be possible until
accurate inelastic cross sections are available for low energies,
requiring quantal calculations and an accurate potential surface.

\subsection{Models\label{sec:models}}

A PDR has hundreds of fluorescent emission lines which may be observable,
but space limitations preclude discussion of other than the
1--0S(1), 2--1S(1), and 6--4Q(1) lines discussed above.
Complete $\HH$ vibration-rotation emission spectra may be obtained 
via anonymous ftp for the models listed in Table \ref{tab:models}.
Readers interested in $\HH$ spectra for other values of
$\chi$, $\nH$, $T_0$, $\sigma_d$, $T_{\rm f}$, 
or $b$ should contact the authors.
\begin{figure}[t]
\epsscale{0.60}%size1
\plotone{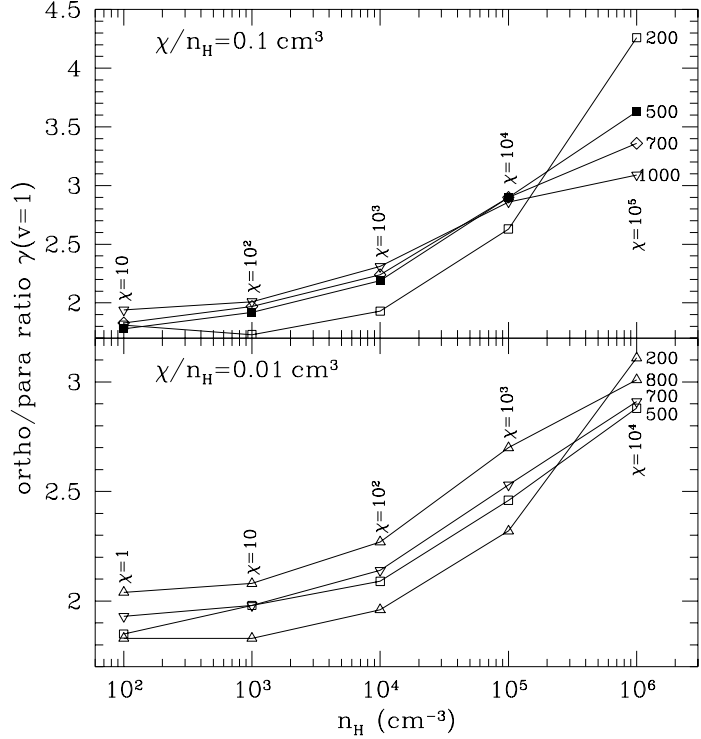}
\caption{
	Same as Fig.~\protect{\ref{fig:10S1/21S1}}, but showing ortho/para
	ratio parameter $\gamma$
	[see eq.\ (\protect\ref{eq:gamma})]
	for $\HH$ in levels $v=1$, $2\leq J\leq7$.
	An $\H^+$ fraction $x_\H=10^{-4}$ has been assumed.
	\label{fig:orthopara}
	}
\end{figure}
\begin{figure}[t]
\epsscale{0.60}%size1
\plotone{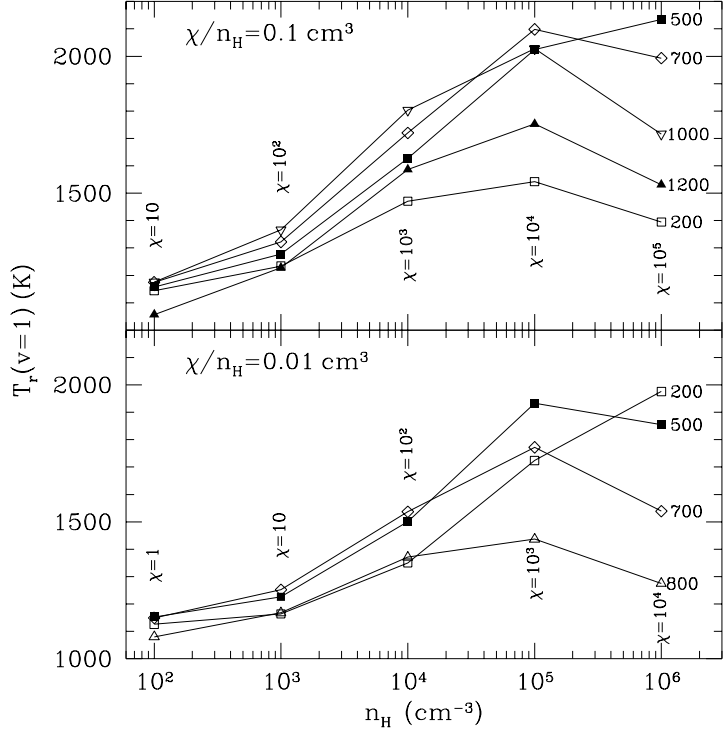}
\caption{
	Same as Fig.~\protect{\ref{fig:10S1/21S1}}, but showing 
	rotational
	temperature of $\HH$ in levels $v=1$, $2\leq J\leq7$.
	\label{fig:trotv=1}
	}
\end{figure}
\begin{figure}[t]
\epsscale{0.60}%size1
\plotone{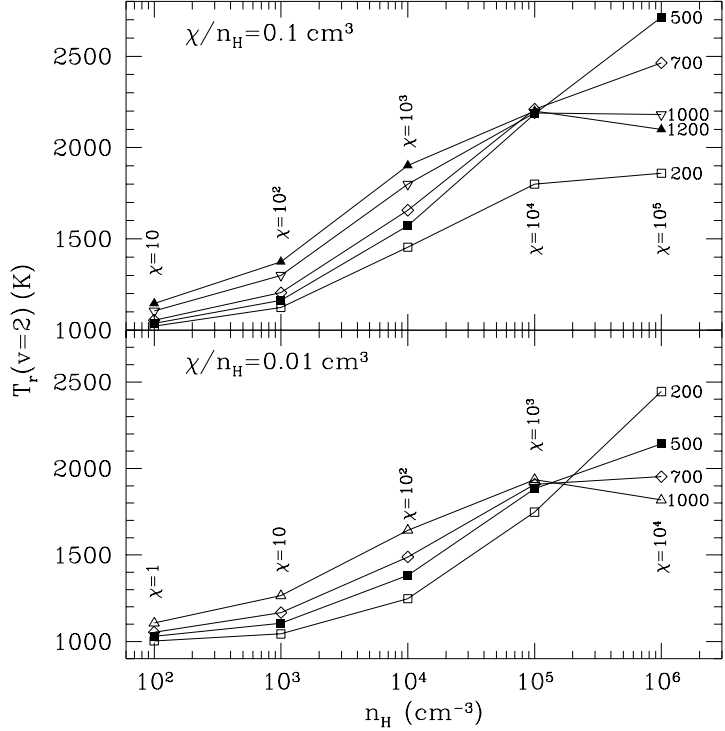}
\caption{
	Same as Fig.~\protect{\ref{fig:trotv=1}}, but for
	levels $v=2$, $2\leq J\leq7$.
	\label{fig:trotv=2}
	}
\end{figure}

\section{Rotational Populations\label{sec:Jdist}}

The rotational populations within a given vibrational level can be
approximated by a thermal distribution,
\beq
N(v,J)/g_J = a \exp(-E(v,J)/kT_{r})~~~,
\label{eq:trot}
\eeq
where
\begin{eqnarray}
g_J &=& (2J+1) ~~~~~{\rm for~even~}J~~(``{\rm para-\HH"}),
\\
&=& 3(2J+1) ~~~~{\rm for~odd~}J~~(``{\rm ortho-}\HH"),
\end{eqnarray}
and $T_r$ is the apparent ``rotational temperature''.
Because ortho- and para-$\HH$ may not be fully equilibrated, we allow
separate values of $a$, denoted $a_{ortho}$ and $a_{para}$,
for the ortho- and para-$\HH$ states, but we fit a single rotational
temperature $T_r$
for both ortho- and para-$\HH$ within a single vibrational level.
We define the ratio
\beq
\gamma\equiv 3a_{ortho}/a_{para}
\label{eq:gamma}
\eeq
to characterize the ortho-para ratio:
from our definition of $\gamma$, we see that
for a thermal distribution we would have $\gamma=3$, even at low
temperatures.
Note that $\gamma$ is {\it not} the actual ortho/para
ratio
(i.e., the ratio of the total column densities
of the ortho--H$_2$ and para--H$_2$), which
goes to zero in thermal equilibrium as $T\rightarrow0$.

In Figure \ref{fig:orthopara} we show $\gamma$ determined by
fitting eq.\ (\ref{eq:trot}) to our model calculations; the fit
is restricted to the $v=1$ levels $2\leq J\leq 7$.
For $\nH\ltsim10^4\cm^{-3}$ and $T_0\ltsim700\K$, we 
find $\gamma\approx2\pm0.2$,
essentially independent of $\chi/\nH$.

As discussed in \S\ref{sec:h2form}, the $\HH$ is assumed to be formed
on grains with an ortho/para ratio of 2.78, so an appreciably 
lower value of $\gamma$
for the $v=1$ levels reflects ortho$\rightarrow$para conversion
in the gas.
We have assumed $x_\H=n(\H^+)/\nH=10^{-4}$ in the present models.
In regions with $x_\H>10^{-7}$,
the dominant ortho-para conversion process is reactive scattering with
$\H^+$, with a rate coefficient $\sim2\times10^{-10}\cm^3\s^{-1}$ for
$\HH(0,1)+\H^+\rightarrow\HH(0,0)+\H^+$.
While this is much slower than photodissociation of $\HH$ in unshielded regions,
note that at the point where $2n(\HH)=n(\H)$ the $\HH$ photodissociation
rate is
$\zeta_{diss}
=R\nH=6\times10^{-17}(T/100\K)^{0.5}(\nH/\cm^{-3})\s^{-1}$,
slow compared to the rate $2\times10^{-14}(\nH/\cm^{-3})(x_\H/10^{-4})\s^{-1}$
for (0,1)$\rightarrow$(0,0) conversion by $\H^+$.
In regions where $x_\H$ is very small, ortho-para conversion takes place
via scattering by H atoms and off grain surfaces.

As $\nH$ is increased above $\sim 10^4\cm^{-3}$, $\gamma$ increases and
approaches $\sim 3$.
Indeed, we see that this non-LTE environment can result in values of
$\gamma>3$: for $\chi=10^5$, $\nH=10^6\cm^{-3}$, and $T_0=200\K$, our
best-fit $\gamma=4.3$.
Note that this occurs under conditions where
UV pumping can compete with spontaneous decay from
vibrationally-excited levels (cf. \S3).
The observed large value of $\gamma$ may result from the greater ability of
ortho-$\HH$ to self-shield.

The rotational temperatures $T_r(v=1)$ and $T_r(v=2)$ of the
$v=1$ and 2 levels of $\HH$ are shown in
Figures \ref{fig:trotv=1} and \ref{fig:trotv=2}.
We see that optical pumping results in rotational temperatures $1000-2000\K$.
There is a general trend for $T_r$ to increase with increasing $\chi$
(e.g., compare lower and upper panels of each Figure), due to the increasing
importance of optical pumping compared to radiative decay or collisional
deexcitation.
It is interesting to note that under some conditions an increase in $T_0$
leads to a decrease in $T_r$: for example, at $\nH=10^4\cm^{-3}$ and
$\chi=10^2$, changing $T_0$ from 500 to 1000K causes $T_r$ to decrease
from 1500 to 1000K.
This is due to the large increase in H-$\HH$ inelastic rate coefficients,
making it possible for collisional processes to compete with radiative
pumping, 
tending to bring $T_r$ toward 
the gas temperature $T$.

\section{NGC 2023}
\subsection{Observations\label{sec:2023observations}}
The B1.5V star HD 37903, at an estimated distance $D=450\pc$, 
is situated near the edge of the
molecular cloud L1630, resulting in the prominent reflection nebula NGC 2023.
NGC 2023 displays strong $\HH$ fluorescent
emission, and its spectrum has been studied both in the
infrared (Gatley \etal 1987; Hasegawa \etal 1987)
and the far-red (Burton \etal 1992).
Gatley \etal have mapped the 1--0S(1) emission, which peaks in an 
``emission ridge'' SSE of the star.
The brightest spot on the ridge is located about 
78\arcsec S, 9\arcsec W of the star
(Burton \etal 1989; Field \etal 1994; Brand 1995),
corresponding to a transverse distance of 
$5.3\times10^{17}(D/450\pc)\cm$.
Field \etal (1994) have obtained a high resolution image of NGC 2023 in the
1--0S(1) line; the brightest spot is 27\arcsec N, 22\arcsec E of the star,
with a surface brightness slightly higher than on the southern emission ridge.
The C recombination line emission peaks in the vicinity of the 
southern 1--0S(1)
emission peak (Wyrowski \& Walmsley 1996); 
C recombination line spectroscopy indicates
a gas density $\nH\approx10^5\cm^{-3}$ (Pankonin \& Walmsley 1976, 1978).
Since the only available far-red spectra of NGC 2023 were taken on the
southern filament, our modelling objective will be to try to reproduce
the observed $\HH$ spectrum of the southern filament.

\begin{figure}[t]
\epsscale{0.60}%size1
\plotone{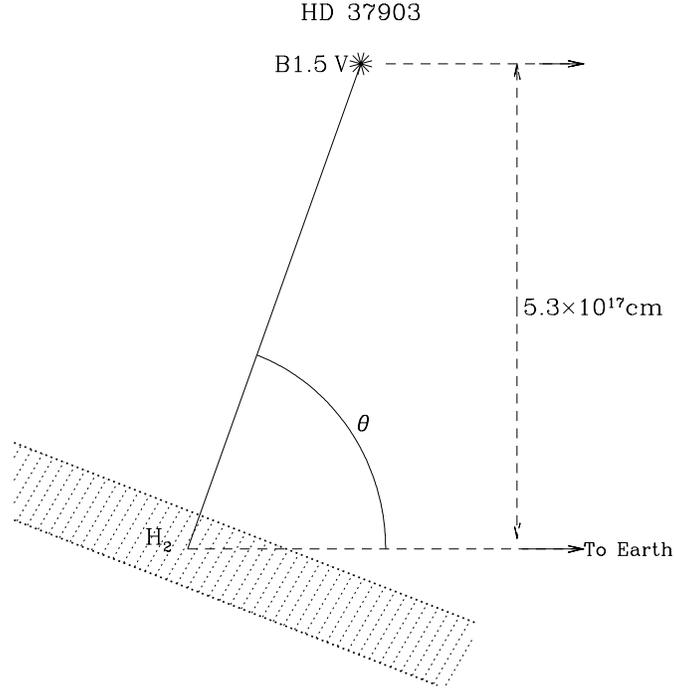}
\caption{
	Slab geometry assumed for the $\HH$ ``emission ridge''
	observed for NGC 2023.
	\label{fig:geometry}
	}
\end{figure}

A B1.5V star ($T_{\rm eff}=22000\K$, $L=7600L_\sol$)
radiates
$S_{uv}\approx 2.5\times10^{47}\s^{-1}$
in the $1110-912\Angstrom$ range
(Fitzpatrick 1995, based on Kurucz ATLAS9 model atmospheres);
at a distance $r$
from the star,
\beq
\chi\approx 
{S_{uv}/4\pi r^2
\over
1.209\times10^7\cm^{-2}\s^{-1}}
=
6600\left(5\times10^{17}\cm\over r\right)^2~~~,
\eeq
neglecting dust extinction.
Harvey \etal conclude that most of
the infrared-emitting dust is more than 0.1pc from HD 37903.
We therefore assume that there is only modest absorption by dust
between the star and the $\HH$ emission ridge,
and seek to reproduce the observed $\HH$ line spectra
by a model of a plane-parallel stationary photodissociation region
illuminated by radiation with $\chi\approx5000$.
The idealized geometry is shown in Figure \ref{fig:geometry}.

The computed surface brightnesses will of course depend on the
adopted value of the angle $\theta$.
Since the highest
fluorescent surface brightness will presumably
be in a region where there is significant limb-brightening, we need to
estimate the likely magnitude of this effect.
A shell with
inner and outer radii $r_i$ and $r_i+\Delta r$, containing material with
emissivity $j$, would have a
peak surface brightness (in the absence of
internal extinction)
$j(8r_i\Delta r)^{1/2}[1+\Delta r/2r_i]^{1/2}$.
Viewed face-on, a portion of the shell would have surface
brightness $j\Delta r_i$, so that limb-brightening enhances the
surface brightness over that of a plane-parallel slab viewed normally
by a factor
$B=(8r_i/\Delta r)^{1/2}[1+\Delta r/2r_i]^{1/2}$.
For $\nH\approx10^5n_5\cm^{-3}$ and 
UV extinction cross section $\sigma_{1000}\approx6\times10^{-22}\cm^2$,
we would estimate the thickness of the fluorescing region to be
$\Delta r\approx (n_{\rm H}\sigma_{1000})^{-1}\approx
1.7\times10^{16}n_5^{-1}\cm$.
Taking $r_i\approx5\times10^{17}\cm$
we then estimate a peak geometric limb-brightening factor
$B\approx 15n_5^{1/2}$.
We regard this as an upper limit to the value of $(\cos\theta)^{-1}$.
If the local radius of curvature of the PDR at the surface of a
``filament'' is smaller than the distance to the star, the
peak limb-brightening factor will be smaller than this upper limit.

A convenient way to compare with observations is to plot
$N_{obs}(v,J)/g_J$ vs. $E(v,J)$, where 
$E(v,J)$ is the energy and
$g_J$ is the degeneracy
of the $(v,J)$ level.
The ``observed'' $N_{obs}(v,J)$ values are obtained from
eq.\ (\ref{eq:nobs}).\footnote{
	Note that the column densities $N(v,J)$ (uncorrected for extinction)
	reported by Burton \etal (1992) are not
	consistent with their reported intensities, due to a
	multiplicative error (Burton 1995);
	the Burton \etal column densities must be reduced
	by a factor 0.36 to
	bring them into agreement with the reported intensities.
	}
We characterize the extinction by $A_K$, and have used the 
Cardelli, Clayton, \&
Mathis (1989) extinction curve for $R_V=5.5$ to 
estimate $A_\lambda/A_K$.
Burton (1993) has estimated the K-band extinction to be $A_K=0.3\mag$;
we assume a slightly smaller extinction $A_K=0.2\mag$, which corresponds
to extinction by dust
with $E(B-V)\approx1.37A_K=0.27\mag$ and $A_V=1.5\mag$.
The extinction at 0.9 and 0.7 $\mu$m is estimated to be
$A_{0.9}=4.2A_K=0.84\mag$\ and $A_{0.7}=6.0A_K=1.2\mag$.
\begin{figure}[t]
\epsscale{0.58}%size3
\plotone{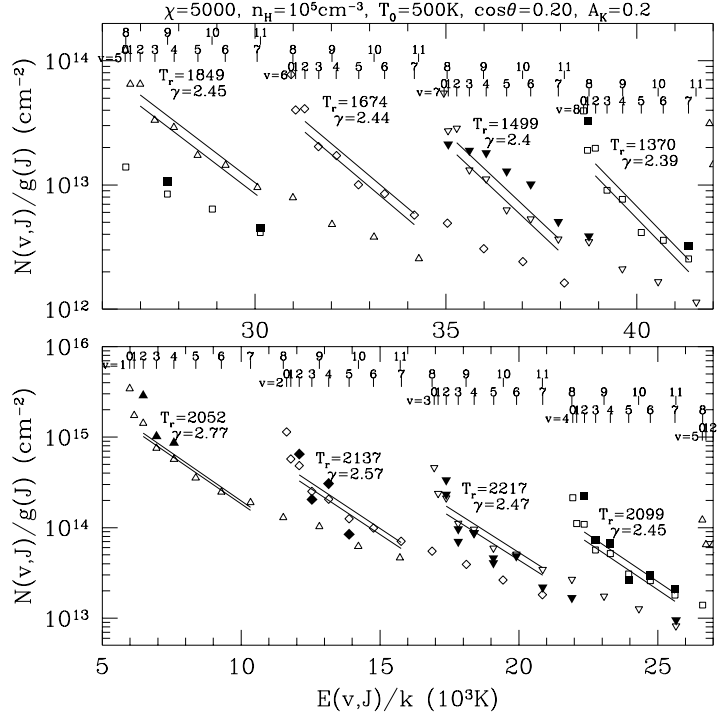}
\caption{
	Column densities $N(v,J)$ divided by the level degeneracy
	$g(J)$ for the NGC 2023 emission ridge.
	Filled symbols are derived from observations of 
	Hasegawa \etal (1987) and Burton et al.(1992), corrected for 
	extinction with $A_K=0.2\mag$.
	Surface brightnesses reported by Hasegawa \etal (1987) have
	been multiplied by 6 to allow for beam dilution (see text);
	surface brightnesses of Burton et al. have been reduced by a 
	factor 0.4 for consistency (see text).
	Open symbols are for a plane-parallel slab model with
	$\chi=5000$, $\nH=10^5\cm^{-3}$, and $T_0=500\K$, 
	viewed at an angle with $\cos\theta=0.2$.
	Solid lines are least-squares LTE fits to the $2\leq J\leq 7$
	populations in each vibrational level.
	For each vibrational level the best-fit values of the rotational
	temperature $T_r$ and ortho/para ratio parameter $\gamma$
	are indicated.
	\label{fig:NvsE_modA}
	}
\end{figure}

	 \begin{figure}[t]
\epsscale{0.60}%size1
\plotone{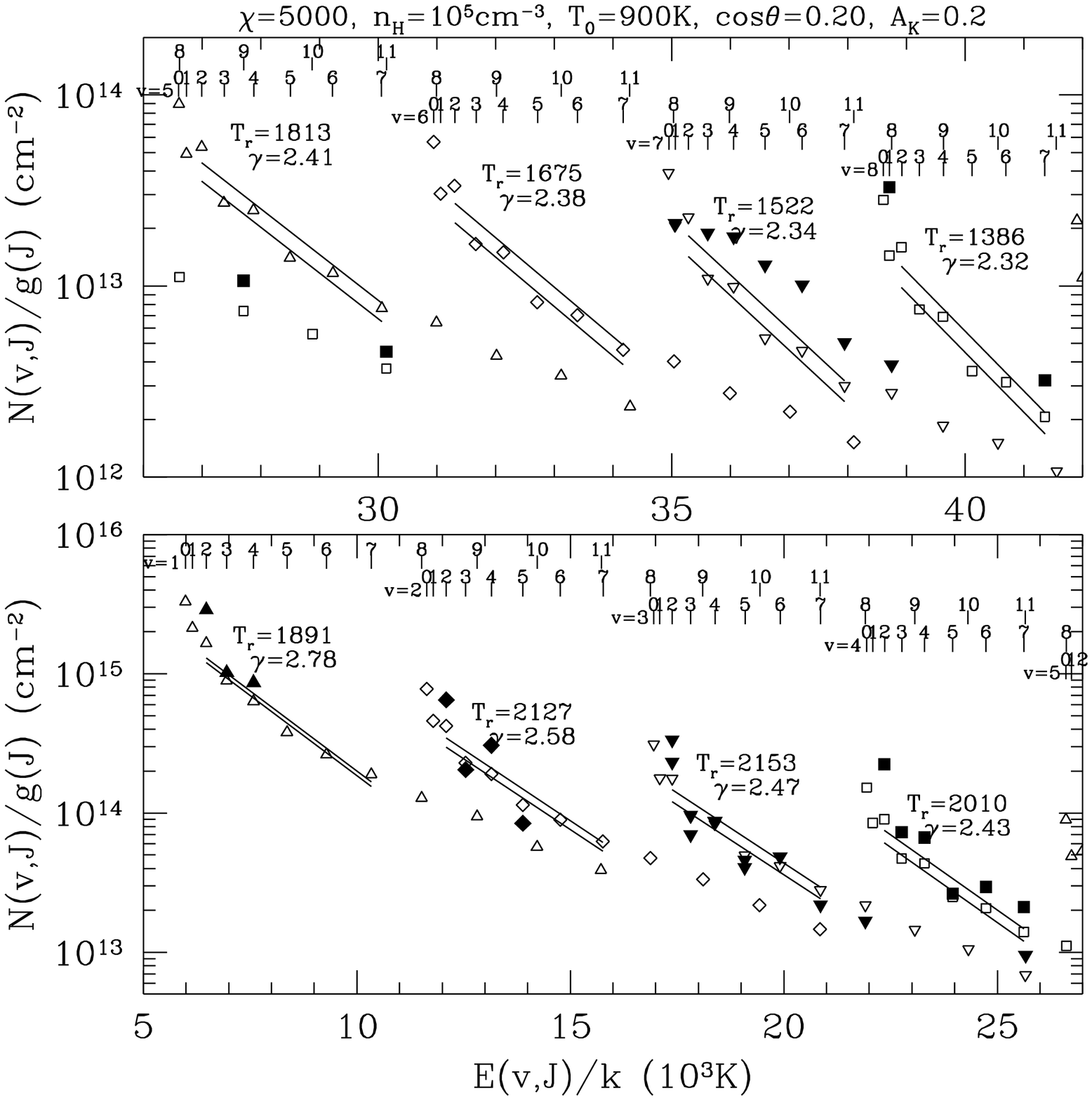}
\caption{
	Same as Figure \protect{\ref{fig:NvsE_modA}}, but for model B, with 
	$\chi=5000$, $\nH=10^5\cm^{-3}$, $T_0=900\K$, and 
	$\cos\theta=0.2$.
	\label{fig:NvsE_modB}
	}
\end{figure}

There are 3 levels ($v\!=\!3, J\!=\!3,4,5$) for which emission has been
reported by both Hasegawa et al. (1987; hereafter H87) and
Burton et al. (1992; hereafter B92).
Unfortunately, after correcting for differential extinction,
the column densities $N_{obs}(v,J)$
derived from the B92 far-red spectrum
are a factor of $\sim10-20$ larger than those derived from the K-band
observations of H87!
For example, the 3--2S(3) flux observed by H87, corrected
for reddening by $A_K=0.2\mag$, 
corresponds to $N_{obs}(3,5)=2.5\times10^{14}\cm^{-2}$.
However, the 3--0S(3) flux reported by B92, corrected for
reddening by $A_{.7962\micron}=5.08A_K=1.02\mag$, corresponds to
$N_{obs}(3,5)=3.4\times10^{15}\cm^{-2}$ --  a factor of 13 larger than
inferred from the K band spectra!  Discrepancies of 21 and 15 are found for the
(3,3) and (3,4) levels, although for these levels the reported uncertainties
in the fluxes are greater.
We will assume an overall discrepancy of 15 between the H87 and B92 column
densities.
The fluorescent emission is strongly peaked on the ``emission ridge''
(Burton et al. 1989, Brand 1995), and the differences between H87 and B92 are 
in large part due to the fact that
H87 used a $19\farcs6$ diameter beam while B92 used
a $27\arcsec\times1\arcsec$ aperture aligned along the
emission ridge.
We will assume that the H87 surface brightnesses need to be multiplied by a 
factor $\beta_{\rm H87}=6$
to correct for beam dilution, but that the B92 fluxes should be multiplied
by a factor $\beta_{\rm B92}=\beta_{\rm H87}/15=0.40$ to correct 
for an apparent
calibration error.\footnote{
	Note that with the H87 intensities increased by a factor of 6,
	the B92 intensities would not require correction if there were no
	differential extinction between the K band and the far-red.
	This possibility does not seem likely, however, as even the
	dust within the PDR should contribute extinction with an effective
	$A_K\approx0.2$.
	}

\subsection{Models for the NGC 2023 PDR \label{sec:2023models}}
We consider two stationary PDR models for the NGC 2023 emission ridge.
Both have $\chi=5000$, $\nH=1\times10^5\cm^{-3}$, and a viewing angle
$\theta=\cos^{-1}0.2\approx78\arcdeg$.
We assume the temperature profile in the PDR to be 
given by eq.\ (\ref{eq:tprof})
and consider two values of $T_0$: model A has $T_0=500\K$, and
model B has $T_0=900\K$.
Both models have $\chi/\nH=0.05\cm^3$;
$\phi_0=0.12$ for model A and 0.086 for model B.
From Fig.\ \ref{fig:Nvj13}, for both models we expect most of the
vibrationally-excited $\HH$ to be within the first
$\NH=1.5\times10^{21}\cm^{-2}$, or $1.5\times10^{16}\cm$, corresponding to
$\sim2$\arcsec~ at the distance of NGC 2023.
This is consistent with the observed narrowness of the ``emission ridge''.

The pumping efficiency $\epsilon_{pump}=0.69$ for model A and 0.74 for model B,
somewhat larger than
the estimates $\sim$0.45 from eq.\ (\ref{eq:eps_pump_approx});
the enhancement of $\epsilon_{pump}$ results from the substantial populations
of $\HH$ in
excited rovibrational levels, so that UV absorption by $\HH$ is
spread over more lines, including many with $\lambda>1110\Angstrom$.
The models succeed in reproducing most of the observed intensities to within 
a factor $\sim 2$.
As we shall see, model B appears to be in fairly good agreement with the
published observations for NGC 2023.

\subsection{1--0S(1)/2--1S(1) Ratio}
Hasegawa et al. find 1--0S(1)/2--1S(1) = $3.7\pm0.6$;
from Figure 2 of Burton (1993) we infer a value of 3.0 for this ratio.
These are both well above the calculated value for our model A, which has
1--0S(1)/2--1S(1)=2.20, a value typical of UV pumping.
Model B, with a higher gas temperature, has 1--0S(1)/2--1S(1)=2.91; this is
in part due to collisional excitation of the $v=1$ level, and in part
to preferential collisional deexcitation of UV-pumped $v=2$,
as discussed in \S\ref{sec:Jdist}.
Varying $T_0$ between 800K and 1000K allows us to obtain
1--0S(1)/2--1S(1) ratios between 2.54 and 3.90.
We consider Model B to be in good agreement with the observed 1--0S(1)/2--1S(1)
ratio.

\subsection{Rotational Temperatures and Ortho-Para Ratio}

In Figures \ref{fig:NvsE_modA} and \ref{fig:NvsE_modB} one sees that
there is fairly good overall agreement between predicted and observed
line intensities.
Figures \ref{fig:NvsE_modA} and \ref{fig:NvsE_modB} show the
least-squares fit to the rotational temperature $T_r$ characterizing
rotational levels $2\leq J\leq7$ for each vibrational level.
H87 obtained a rotational temperature $T_r=900\K$ for the $v=1$,
$J=2-4$ levels.  If we restrict our least-squares fit to these 3 levels,
we obtain $T_r=1217\K$ and $\gamma=2.36$ for Model A and 
$T_r=1160\K$, $\gamma=2.46$ for Model B, in 
reasonable agreement with H87.

H87 found $T_r\approx1500\pm200\K$ for the $v=2$, $J=2-5$ levels.
For these 4 levels, our least-squares fit gives
$T_r=1587\K$ for model A, and $T_r=1650\K$ for model B.

For the higher vibrational levels we also do a fairly good job
of reproducing the relative populations of the different J levels,
as can be seen for the $v=3$, 4, 7, and 8 levels in Figures
\ref{fig:NvsE_modA} and \ref{fig:NvsE_modB}.

Our ortho-para ratios are somewhat above the observed values.
Burton (1993) finds that the observed line intensities are consistent
with $\gamma=2.1\pm0.2$ for all levels, whereas we find (for Model B)
$\gamma=2.46$ for $v=1$ ($J=2-4$), and $\gamma=2.46$ for $v=2$ ($J=2-5$).
\begin{figure}[t]
\epsscale{0.56}%size4
\plotone{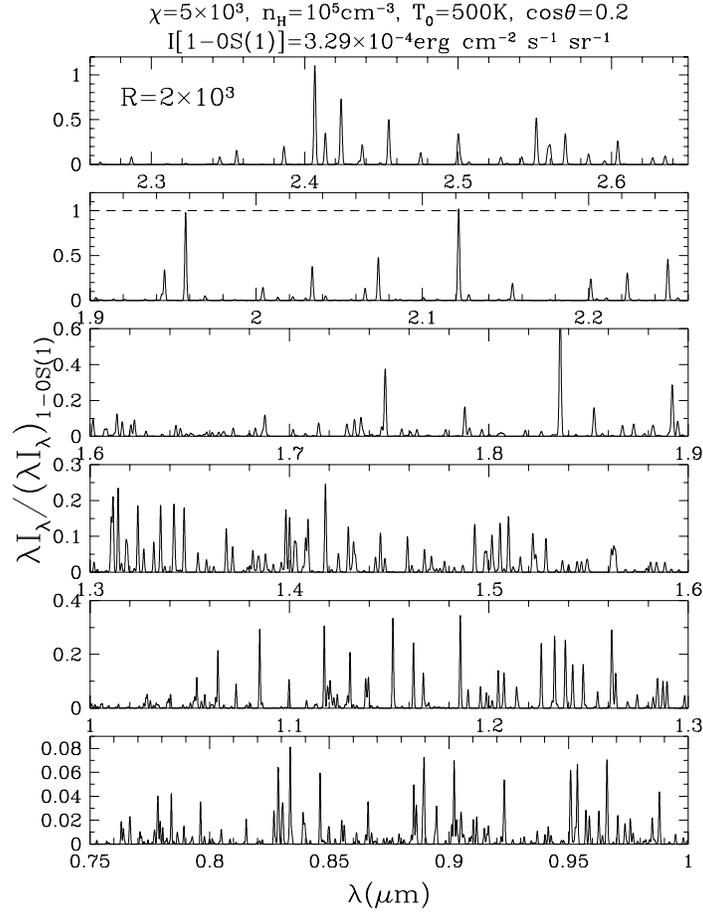}
\caption{
	Emission spectrum for NGC 2023 model A ($\chi=5000$,
	$\nH=10^5\cm^{-3}$, $T_0=500\K$) viewed from a direction with
	$\cos\theta=0.2$.
	The spectrum is convolved with a Gaussian
	response function with
	$R=\lambda/{\rm FWHM}_\lambda=2\times10^3$.
	Extinction by dust within the PDR has been included, but no external
	dust has been allowed for.
	\label{fig:spec_modA}
	}
\end{figure}
\begin{figure}[t]
\epsscale{0.60}%size1
\plotone{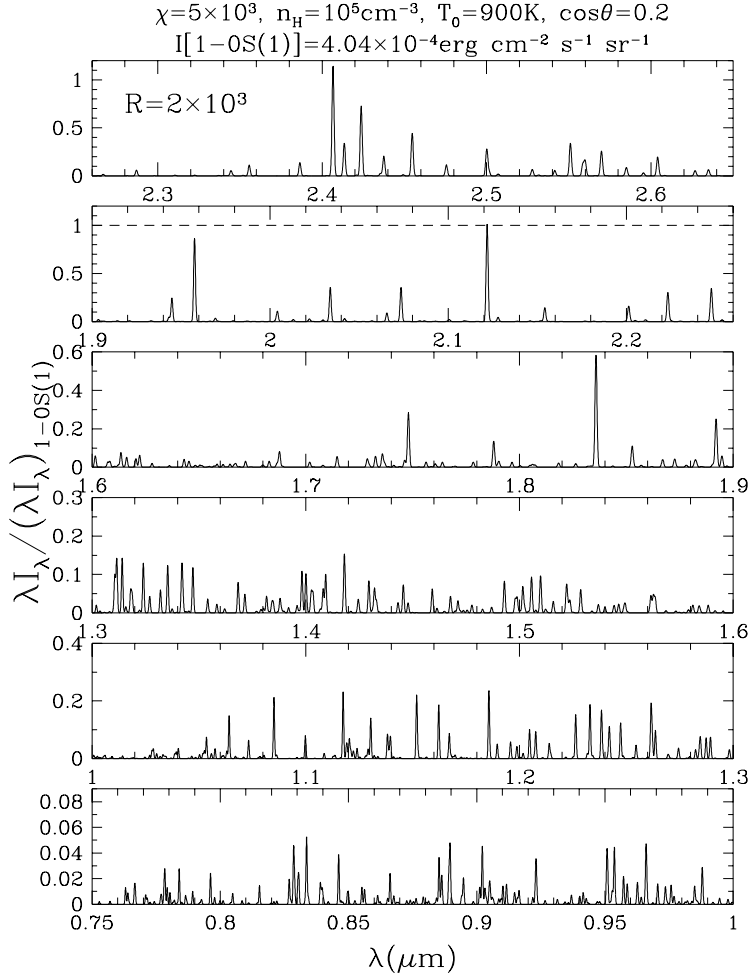}
\caption{
	Same as Fig.\ \protect{\ref{fig:spec_modA}}, 
	but for model B with $T_0=900\K$.
	\label{fig:spec_modB}
	}
\end{figure}

\subsection{Spectra}

We have taken the predicted emission spectra for the above models and
convolved them with a Gaussian
with 
$\lambda/{\rm FWHM}_\lambda=2000$ 
(FWHM$_v=150\kms$).
The spectra include the effects of
extinction within the emitting slab,
but no additional foreground extinction
is allowed for.
In Figures \ref{fig:spec_modA} and 
\ref{fig:spec_modB} we show the predicted emission spectra for models A and B.
The far-red lines derive from transitions from high-$v$ states, for which
the collisional deexcitation rate coefficients are much larger than for the
low-$v$ states.
Note that the far-red emission is relatively stronger in model A than in
model B -- this is because there is more collisional deexcitation in model B
due to the higher temperature.
A list of all the $\HH$ lines with intensities exceeding 0.1% of 1--0S(1)
(1280 and 1174 lines for models A and B, respectively) is available
via anonymous ftp (see Table \ref{tab:models}).

\subsection{Discussion}

Black \& van Dishoeck (1987) have previously put forward
models for the fluorescent emission from NGC 2023.
Their models had $I_{UV}=300$ which
(cf. Table 1)
corresponds to $\chi\approx670$ -- a factor of 7.5 weaker
than the $\chi=5000$ favored here.
The reported 1--0S(1) {\it face-on}
surface brightness for their ``reference model A'' is
$3.5\times 10^{-5}\erg\cm^{-2}\s^{-1}\sr^{-1}$,
corresponding to an ``efficiency''
$\epsilon_{1-0S(1)}=.058$.
We have computed a model with $\chi=670$, $\nH=10^4\cm^{-3}$,
$T=80\K$, and the same grain properties (extinction and $\HH$ formation
rate) as assumed by Black \& van Dishoeck for their reference model A;
we obtain an efficiency $\epsilon_{1-0S(1)}=0.021$, a factor of
2.7 smaller than obtained by Black \& van Dishoeck.
Part of the discrepancy is attributable to the fact that Black \& van Dishoeck
did not allow for the effects of line overlap; if we modify our model to
neglect line overlap (by taking $W_{max}\rightarrow\infty$ in
eq.\ \ref{eq:withoverlap}), the 1--0S(1) surface brightness increases by
a factor of 1.5.
The remaining discrepancy, amounting to a factor of $2.7/1.5=1.8$, 
may be due in part to the fact that we use improved (larger) 
dissociation probabilities for the $C^+$ levels, although
this seems unlikely to explain the factor of 1.8.

The Black \& van Dishoeck models were developed with the aim of reproducing
the surface brightnesses reported by H87.
As discussed in \S\ref{sec:2023observations}, we now believe the
peak 1--0S(1) surface brightness to be a factor $\sim6$ higher than
reported by H87, 
so the published Black \& van Dishoeck
models are not applicable to
the bright bar in NGC 2023.

Our modelling efforts lead us to conclude that if the dissociation front
is approximately stationary, then the fluorescing $\HH$ in
the NGC 2023 emission ridge must have $\nH\approx10^5\cm^{-3}$, and appreciable
limb-brightening (such as $\cos\theta=0.2$, as in our Models A and B),
in order to reproduce the observed high surface brightness.
Lowering $\nH$ (increasing $\chi/\nH$) would lead to somewhat lowered 
fluorescent efficiency,
and by reducing the importance of collisional effects would make it
difficult to reproduce the observed 1--0S(1)/2--1S(1) line ratio,
which requires a combination of collisional deexcitation from $v=2$,
and collisional excitation from $v=0$ to 1.
On the other hand, increasing $\nH$ would result in increased collisional
deexcitation of vibrationally-excited levels, making it difficult to
reproduce the observed high surface brightnesses.
It is reassuring that our density estimate $\nH\approx10^5\cm^{-3}$ is
in agreement with that deduced by Pankonin \& Walmsley (1976, 1978)
from C recombination lines, which should originate from the same gas.
Jansen, van Dishoeck, \& Black (1994) have used the observed
HCO$^+$3--2/4--3 line ratio to infer
$\nH\approx5\times10^4\cm^{-3}$ in this region.
Fuente, Martin-Pintado, and Gaume (1995) have argued for 
$\nH\approx10^6\cm^{-3}$ near the photodissociation front based on
observations of CN, SO, and HCN rotational lines.

Our best model (B) does not fully reproduce the observed line surface
brightnesses (see Fig. \ref{fig:NvsE_modB}).
There are a number of possible explanations:
(1) The emission bar may include a range of densities, whereas in our
modelling we seek to explain the observations with a single density.
(2) The actual temperature profile may be rather different from that
assumed here.
(3) Line overlap effects will suppress particular photoexcitation
transitions, whereas in our treatment we share this suppression among 
all of the transitions with $1110<\lambda<912\Angstrom$.
(4) As discussed in \S\ref{sec:radrates}, the actual branching ratios following
photoexcitation in the damping wings may differ from our adopted branching
ratios, thereby altering relative line strengths.
(5) Errors in our adopted collisional rate coefficients may be important,
since the vibrationally-excited states may be collisionally-deexcited at
the densities $\nH\approx10^5\cm^{-3}$ assumed here.
(6) The dissociation front in NGC 2023 may be propagating rapidly enough
that the stationary approximation made here leads to errors.

NGC 2023 shows structure on $\sim$1\arcsec~scales (Field \etal 1994).
We need calibrated $\HH$ spectral-line images of NGC 2023 in a number
of $\HH$ fluorescence lines to test the accuracy of our models.

\section{Summary}

The principal results of this paper are as follows:
\begin{enumerate}
\item An approximate treatment of self-shielding [eq.\ (\ref{eq:withoverlap})]
is developed
which treats line overlap in a statistical fashion.
This treatment allows individual lines to be treated using the single-line
equivalent-width approximation, but allows statistically
for overall suppression of the continuum due to line overlap.
Comparison with exact calculation shows this approximation to be accurate.
\item Two simple approximations are provided for the self-shielding function
for $\HH$.  The first, a simple power-law [eq.\ (\ref{eq:powfit})], provides a 
fairly good fit to self-shielding over the range 
$10^{15}< N_2 < 10^{21}\cm^{-2}$.
The second [eq.\ (\ref{eq:goodfit})] is an analytic function which provides 
a very good approximation to the self-shielding including the effects 
of line overlap.
These functions are recommended for use in future studies of photodissociation
regions.
\item The effects of line overlap become
important for column densities $N(\HH)\gtsim10^{20}\cm^{-2}$,
suppressing the UV pumping rate by a factor of 2 at
$N(\HH)\approx3\times10^{20}\cm^{-2}$
(cf. Fig.~\ref{fig:eqwidths}).
\item For dense cloud dust properties,
the effects of line overlap become important
while the dust is
still optically thin for
$\chi/\nH\ltsim 0.05$ (cf. Fig.~\ref{fig:NH2vsNH,dense}).
\item The dust optical depth $\tau_{pdr}$ 
at the point where $2n(\HH)=n({\rm H})$ is
primarily a function of a single parameter 
$\phi_0 \propto (\chi/\nH R)\sigma_{1000}^{3/4}$, defined in 
eq.\ (\ref{eq:phi_0}).
The approximation (\ref{eq:taupdr}) provides a good estimate for
$\tau_{pdr}$.
\item The ``efficiency'' of UV pumping of $\HH$
in the photodissociation front,
$\epsilon_{pump}$,
is also a function of the single parameter $\phi_0$,
and is approximately given by eq.\ (\ref{eq:eps_pump_approx}).
\item For $\nH\ltsim10^5\cm^{-3}$ the efficiency
$\epsilon_{1-0S(1)}$ for 1--0S(1) emission is given as a function of $\phi_0$
and $\chi$ by eq.\ (\ref{eq:epsilon10s1fit}).
\item The emission from PDRs is quite insensitive to
the color temperature of the illuminating radiation for color
temperatures $T_{color}\gtsim10^4\K$, or stars with spectral
type A0 or earlier.
\item The emission spectrum from low $J$ levels [e.g., 1--0S(1), or 3--2S(1)]
is insensitive to the distribution function $\hhini(v,J)$ of
newly-formed $\HH$, but the emission from higher $J$ levels [e.g., 1--0S(9)]
is increased
(by a factor $\sim2$) when the newly-formed $\HH$ is rotationally-``hot''.
\item Observable properties of photodissociation fronts -- including
the 1--0S(1)/2--1S(1) line ratio (Fig.~\ref{fig:10S1/21S1}), 
the 2--1S(1)/6--4Q(1) line ratio (Fig.~\ref{fig:21S1/64Q1}),
the ortho/para ratio (Fig.~\ref{fig:orthopara}), and
the rotational temperatures (Fig.~\ref{fig:trotv=1},\ref{fig:trotv=2}) --
are computed for various values of $\chi$, $\nH$, and temperature.
Complete $\HH$ vibration-rotation spectra are available via anonymous ftp.
\item The reflection nebula NGC 2023 is considered.
We correct the H87 intensities for assumed beam dilution, and adjust
the B92 intensities for consistency.
The (adjusted) observations are approximately reproduced by 
a model with $\nH=10^5\cm^{-3}$ and $\chi=5000$, with $T_0\approx900\K$.
The agreement is not perfect; possible explanations for the discrepancies
are discussed.
\end{enumerate}

\acknowledgements

We are especially grateful to 
P.W.J.L. Brand
and M. Walmsley 
for communicating recent observations of NGC 2023 in advance
of publication,
to E. L. Fitzpatrick for computing $S_{uv}$ for B stars,
to R.H. Lupton for making available the SM plotting package,
to P.G. Martin for making available H-$\HH$ rate coefficients in
advance of publication,
and
to E. Roueff for kindly providing us with molecular data
for $\HH$ in computer-readable form.
We wish also to thank 
M.G. Burton,
E. van Dishoeck,
D. Neufeld,
M. Walmsley,
D.T. Jaffe,
and E. Roueff
for helpful comments and discussions.

This research was supported in part by NSF grant AST-9319283, 
and by the Deutsche Forschungsgemeinschaft.

%------------------------------------------------------------------------------
%                           Table 1
\begin{deluxetable}{c c c c c}
\tablecaption{Interstellar Ultraviolet Radiation Fields ($\lambda > 912\Angstrom$)}
\tablehead{
\colhead{Radiation Field}&
\colhead{$\chi$}&
\colhead{$F/\chi$}&
\colhead{$d\ln u_\nu/d\ln\nu$}&
\colhead{$T_{color}$}
	\\
\colhead{}&
\colhead{}&
\colhead{($\cm^{-2}\s^{-1}$)}&
\colhead{(at $1000\Angstrom$)}&
\colhead{($10^4$K)}
	}
\startdata
$u_\nu\!=\!2.84\!\times\!10^{-18}(4\pi/c)B_\nu(4\!\times\!10^4\K)$&
	1&	$1.186\times10^7$&	$-0.302$&	$4.00$\nl
$u_\nu\propto\nu^{-1}$&
	&	$1.195\times10^7$&	$-1$&		$3.67$\nl
$u_\nu\!=\!3.04\!\times\!10^{-19}(4\pi/c)B_\nu(3\!\times\!10^4\K)$&
	1&	$1.197\times10^7$&	$-1.836$&	$3.00$\nl
$u_\nu\propto\nu^{-2}$&
	&	$1.208\times10^7$&	$-2$&		$2.90$\nl
Habing (1968)[eq.\protect{\ref{eq:habing}}]&
	1&	$1.222\times10^7$&	$-3.043$&	$2.39$\nl
$u_\nu\!=\!1.06\!\times\!10^{-16}(4\pi/c)B_\nu(2\!\times\!10^4\K)$&
	1&	$1.237\times10^7$&	$-4.199$&	$2.00$\nl
Mathis, Mezger \& Panagia (1983)&
	1.23&	$1.232\times10^7$&	$-5.417$&	$1.71$\nl
Black \& van Dishoeck (1987)&
	$2.23I_{UV}
	$&	$1.218\times10^7$&	$-6$&	$1.60$\nl
Draine (1978)[eq.\protect{\ref{eq:draine78}}]&
	1.71&	$1.232\times10^7$&	$-8.119$&	$1.29$\nl
$u_\nu\!=\!1.42\!\times\!10^{-13}(4\pi W/c)B_\nu(10^4\K)$&
	1&	$1.510\times10^7$&	$-11.39$&	$1.00$\nl
\enddata
\end{deluxetable}
%                      end of Table 1
%------------------------------------------------------------------------------
%-------------------------------------------------------------------------------
%                       begin Table 2
\begin{deluxetable}{c c c c c c c}
\tablecaption{Pumping and Dissociation Rates for Unshielded $\HH$}
\tablehead{
\colhead{$\HH$ level}
&\multicolumn{3}{c}{\underline{
\quad\quad\quad$u_\lambda$ from eq.(\ref{eq:u_lambda}), $\chi=1$
\quad\quad\quad}}
&\multicolumn{3}{c}{\underline{
\quad\quad\quad$u_\lambda$ from eq.(\ref{eq:draine78}), $\chi=1$
\quad\quad\quad}}
\\
\colhead{$(v,J)$}
&\colhead{$\zeta_{pump}(\s^{-1})$}
&\colhead{$\langle p_{diss}\rangle$}
&\colhead{$\langle p_{ret}\rangle$}
&\colhead{$\zeta_{pump}(\s^{-1})$}
&\colhead{$\langle p_{diss}\rangle$}
&\colhead{$\langle p_{ret}\rangle$}
	}
\startdata
 (0,0)&3.08(-10)&0.134&0.077&2.77(-10)&0.117&0.071\nl
 (0,1)&3.09(-10)&0.136&0.101&2.79(-10)&0.119&0.097 \nl
 (0,2)&3.13(-10)&0.135&0.096&2.84(-10)&0.119&0.091 \nl
 (0,3)&3.15(-10)&0.145&0.094&2.91(-10)&0.129&0.089 \nl
 (0,4)&3.21(-10)&0.154&0.096&3.00(-10)&0.139&0.092 \nl
 (0,5)&3.26(-10)&0.155&0.098&3.11(-10)&0.141&0.093 \nl
 (0,6)&3.38(-10)&0.169&0.098&3.27(-10)&0.157&0.094 \nl
 (0,7)&3.47(-10)&0.170&0.098&3.44(-10)&0.160&0.096 \nl
 (0,8)&3.57(-10)&0.185&0.096&3.63(-10)&0.175&0.093 \nl
 (0,9)&3.71(-10)&0.202&0.096&3.86(-10)&0.195&0.094 \nl
 (1,0)&3.52(-10)&0.083&0.080&4.19(-10)&0.051&0.083 \nl
 (1,1)&3.60(-10)&0.090&0.109&4.23(-10)&0.055&0.114 \nl
 (1,2)&3.64(-10)&0.091&0.101&4.29(-10)&0.056&0.106 \nl
 (1,3)&3.68(-10)&0.091&0.100&4.36(-10)&0.057&0.104 \nl
 (1,4)&3.73(-10)&0.096&0.100&4.46(-10)&0.062&0.104 \nl
 (1,5)&3.81(-10)&0.108&0.098&4.58(-10)&0.075&0.102 \nl
 (2,0)&4.23(-10)&0.111&0.081&5.67(-10)&0.078&0.084 \nl
 (2,1)&4.25(-10)&0.106&0.110&5.70(-10)&0.074&0.116 \nl
 (2,2)&4.30(-10)&0.109&0.101&5.76(-10)&0.076&0.107 \nl
 (2,3)&4.35(-10)&0.117&0.099&5.84(-10)&0.083&0.104 \nl
 (2,4)&4.39(-10)&0.115&0.097&5.94(-10)&0.084&0.102 \nl
 (2,5)&4.44(-10)&0.126&0.101&6.06(-10)&0.095&0.106 \nl
$T_r=50$&3.08(-10)&0.134&--&2.77(-10)&0.117&--\nl
$T_r=100$&3.09(-10)&0.135&--&2.78(-10)&0.119&--\nl
$T_r=200$&3.09(-10)&0.136&--&2.79(-10)&0.119&--\nl
\enddata
\end{deluxetable}
%                    end Table 2
%-------------------------------------------------------------------------------
%-------------------------------------------------------------------------------
%                   begin Table 3
\begin{deluxetable}{c c c c c c c}
\tablecaption{Library of Stationary PDR Models\tablenotemark{a}
	\label{tab:models}
	}
\tablehead{
\colhead{model}&
\colhead{$n_{\rm H}$}&	
\colhead{$\chi$
	\tablenotemark{b}}&
\colhead{$\sigma_{d,1000}$}&
\colhead{$T_0$
	\tablenotemark{c}}&
\colhead{$\cos\theta$ 
	\tablenotemark{d}}&
\colhead{$I(1-0S(1))$}
	\\
\colhead{}&
\colhead{(${\rm cm}^{-3}$)}&
\colhead{}&
\colhead{$10^{-22}{\rm cm}^2$}&
\colhead{(K)}&
\colhead{}&
\colhead{${\rm erg}\,{\rm cm}^{-2}\,{\rm sr}^{-1}\,{\rm s}^{-1}$}
	}
\startdata
am3d	&$10^2$	&$1$	&20	&300&	1&	$2.16\times10^{-8}$\nl
am3o	&$10^2$	&$1$	&6	&300&	1&	$3.38\times10^{-8}$\nl
aw3d	&$10^2$	&$1$	&20	&500&	1&	$2.41\times10^{-8}$\nl
aw3o	&$10^2$	&$1$	&6	&500&	1&	$3.70\times10^{-8}$\nl
bw3d	&$10^2$	&$10$	&20	&500&	1&	$1.18\times10^{-7}$\nl
bw3o	&$10^2$	&$10$	&6	&500&	1&	$2.36\times10^{-7}$\nl
bh3d	&$10^2$	&$10$	&20	&1000&	1&	$1.50\times10^{-7}$\nl
bh3o	&$10^2$	&$10$	&6	&1000&	1&	$3.30\times10^{-7}$\nl
Bm3o	&$10^3$	&$10$	&6	&300&	1&	$3.54\times10^{-7}$\nl
Bw3o	&$10^3$	&$10$	&6	&500&	1&	$3.92\times10^{-7}$\nl
Cw3o	&$10^3$	&$10^2$	&6	&500&	1&	$2.36\times10^{-6}$\nl
Ch3o	&$10^3$	&$10^2$	&6	&1000&	1&	$3.08\times10^{-6}$\nl
Gm3o	&$10^4$	&$10^2$	&6	&300&	1&	$3.41\times10^{-6}$\nl
Gw3o	&$10^4$	&$10^2$	&6	&500&	1&	$3.65\times10^{-6}$\nl
Hw3o	&$10^4$	&$10^3$	&6	&500&	1&	$1.85\times10^{-5}$\nl
Hh3o	&$10^4$	&$10^3$	&6	&1000&	1&	$2.20\times10^{-5}$\nl
Lm3o	&$10^5$	&$10^3$	&6	&300&	1&	$2.61\times10^{-5}$\nl
Lw3o	&$10^5$	&$10^3$	&6	&500&	1&	$2.69\times10^{-5}$\nl
Mw3o	&$10^5$	&$10^4$	&6	&500&	1&	$9.66\times10^{-5}$\nl
Mh3o	&$10^5$	&$10^4$	&6	&1000&	1&	$1.17\times10^{-4}$\nl
Qm3o	&$10^6$	&$10^4$	&6	&300&	1&	$1.32\times10^{-4}$\nl
Qw3o	&$10^6$	&$10^4$	&6	&500&	1&	$1.42\times10^{-4}$\nl
Rw3o	&$10^6$	&$10^5$	&6	&500&	1&	$3.86\times10^{-4}$\nl
Rh3o	&$10^6$	&$10^5$	&6	&1000&	1&	$4.66\times10^{-4}$\nl
n2023a	&$10^5$	&5000	&6	&500&	0.2&	$3.29\times10^{-4}$\nl
n2023b	&$10^5$	&5000	&6	&900&	0.2&	$4.04\times10^{-4}$\nl
\enddata
\tablenotetext{a}{Models available via anonymous ftp from 
	{\tt astro.princeton.edu},
	subdirectory {\tt draine/pdr}, or at
	{\tt http://www.astro.princeton.edu/}$\sim${\tt draine/} .
	All models have $b=3\,{\rm km}\,{\rm s}^{-1}$,
	newly-formed $\HH$ with $T_{\rm f}=5\times10^4$K,
	and incident UV with $u_\nu\propto\nu^{-2}$.
	}
\tablenotetext{b}{cf. eq.(\protect{\ref{eq:chidef}})}
\tablenotetext{c}{Parameter determining temperature profile via
	eq.(\protect{\ref{eq:tprof}}).}
\tablenotetext{d}{$\theta =$ viewing angle.}
\end{deluxetable}
%
%                         end Table 3
%-------------------------------------------------------------------------------
%	end table of models
\end{document}